\documentclass[journal,10pt]{IEEEtran}
\usepackage{amsfonts}
\IEEEoverridecommandlockouts

\ifCLASSINFOpdf
\else
\fi
\usepackage{epsfig}
\usepackage{graphicx}
\usepackage{psfig}
\usepackage{subfigure}
\usepackage{epsf}
\usepackage[cmex10]{amsmath}
\usepackage{booktabs}
\usepackage{fancyhdr}

\usepackage{color}
\newcommand{\bc}{\begin{center}}
	\newcommand{\ec}{\end{center}}
\newcommand{\be}{\begin{equation}}
\newcommand{\ee}{\end{equation}}
\newcommand{\bea}{\begin{eqnarray}}
\newcommand{\eea}{\end{eqnarray}}

\renewcommand\IEEEkeywordsname{Index Terms}

\begin{document}
	
	\title{Energy Efficient Robust Beamforming and Cooperative Jamming Design for IRS-Assisted MISO Networks}
	\author{Qun~Wang,~\IEEEmembership{Student Member,~IEEE,} Fuhui~Zhou,~\IEEEmembership{Senior Member,~IEEE,}\\ Rose~Qingyang~Hu,~\IEEEmembership{Fellow,~IEEE,}~and~Yi~Qian,~\IEEEmembership{Fellow,~IEEE}
        \thanks{Manuscript received April 30, 2020; revised September 5, 2020 and November 24, 2020; accepted November 26, 2020.  Date of current version November 24, 2020. The project has been sponsored by the National Science Foundation under grants NSF EARS-1547312, EARS-1547330, CNS-2007995, and CNS-2008145. The research of Prof. Zhou was supported by the Natural Science Foundation of China under Grants 62071223, 62031012, 61631020, and 61827801, and Young Elite Scientist Sponsorship Program by CAST. This paper was presented in part at the IEEE International Conference on Communications, Dublin, Ireland, June 2020 \cite{irsqun}. The editor coordinating the review of this article and approving it for publication was Prof. Ya-Feng Liu. \emph {(Corresponding author: Rose Qingyang Hu.)}}
		\thanks{Q.~Wang,~and~R.~Q.~Hu are with the Department
			of Electrical and Computer Engineering, Utah State University, Logan, UT 84322 USA (e-mail: claudqunwang@ieee.org, rose.hu@usu.edu).}
			\thanks{F. Zhou is the College of Electronic and Information Engineering, Nanjing University of Aeronautics and Astronautics, Nanjing, 210000, P. R. China (e-mail: zhoufuhui@ieee.org). F. Zhou is also with Key Laboratory of Dynamic Cognitive System of Electromagnetic Spectrum Space of Nanjing University of Aeronautics and Astronautics.
}
			\thanks{Y.~Qian is with the Department of Electrical and Computer Engineering, University of Nebraska–Lincoln, Omaha, NE 68182 USA (e-mail: yqian2@unl.edu).}
	}
	\maketitle
	
	\IEEEpeerreviewmaketitle
	\begin{abstract}
	Energy-efficient design and secure communications are of crucial importance in  wireless communication networks. However, the energy efficiency achieved by using physical layer security can be limited by the channel conditions. In order to tackle this problem, an intelligent reflecting surface (IRS) assisted multiple input single output (MISO) network with independent cooperative jamming is studied. The energy efficiency is maximized by jointly designing the transmit and jamming beamforming and IRS phase-shift matrix under both the perfect channel state information (CSI) and the imperfect CSI.
	In order to tackle the challenging non-convex fractional problems, an algorithm based on semidefinite programming (SDP) relaxation is proposed for solving energy efficiency maximization problem under the perfect CSI case while an alternate optimization algorithm based on $\mathcal{S}$-procedure is used for solving the problem under the imperfect CSI case.
	Simulation results demonstrate that the  proposed design outperforms the benchmark schemes in term of energy efficiency. Moreover, the tradeoff between energy efficiency and the secrecy rate is found in the IRS-assisted MISO network. Furthermore, it is shown that  IRS can help improve energy efficiency even with the uncertainty of the CSI. 
\end{abstract}
\IEEEpeerreviewmaketitle
\section{Introduction}
\IEEEPARstart{R}{ecently}, the fifth-generation (5G) wireless communication networks have achieved great advancement, and their commercialization is forthcoming \cite{rq1}. Meanwhile, the sixth-generation (6G) wireless networks have been attracting increasing attention from both academia and industry \cite{6g}. They aim for realizing ultra-high spectrum and energy efficiency, ultra-dense user connectivity, and very low latency. Particularly, energy efficiency is of crucial importance in the 6G wireless communication network since diverse energy-intensive communication services are emerging and the reduction of the greenhouse gas emission caused by communication technologies is imperative \cite{cwh1}, \cite{ee1}.   Extensive works have formulated energy efficiency optimization frameworks and designed energy-efficient resource allocation schemes in different networks \cite{hj1}.

Besides energy efficiency, secure communications are also very important in the 6G wireless networks since the communication environments are increasingly complicated and both the security and privacy of user information need to be protected \cite{ee2}. To this end, there exist two major categories for secure communication techniques. One category focuses on traditional cryptographic techniques and the other one is the physical layer security \cite{sc1}. 
Physical layer security has received great attention in recent years. It can achieve secure communications without extra overhead caused by protecting the security key \cite{rq2}. However, the secrecy rate achieved by the mutual information difference between the legitimate user and the eavesdropper is limited as it depends on the difference between the channel condition from the base station to the legitimate user and that from the base station to the eavesdroppers \cite{zc1}. In order to address this issue, many promising techniques have been considered, including artificial noise (AN) and cooperative jammer (CJ) \cite{yw1}.


Lately, due to its potential of simultaneously improving energy efficiency and achieving secure communications,  intelligent reflecting surface (IRS) has attracted significant attention from research community. IRS consists of a large number of low-cost passive reflecting elements with the adjustable phase shifts \cite{irs1}.  By properly adjusting the phase shifts of the IRS’s elements, their reflected signals can combine with those from other paths coherently to enhance the link achievable rate at the receiver and decrease the rate at the eavesdroppers \cite{cwh2}. Moreover, since the IRS does not employ any transmit radio frequency (RF) chains, energy consumption only comes from reflective elements phase adjustment, which is usually very low \cite{irs2}. Thus, IRS is promising to increase the energy efficiency of the wireless communication network and improve the system security \cite{irs4}. Furthermore, it was shown that the secure performance can be significantly improved by cooperative jamming \cite{mcj3}. Thus, it is envisioned that IRS-assisted cooperative jamming is promising to further improve the secrecy rate of the legitimate users. 

Note that most of the existing works that focus on IRS-assisted secure communication networks assumed that the channel state information of the link from the IRS to the eavesdropper can be perfectly obtained \cite{irs3}, \cite{cwh4}. However, in practice, it is extremely difficult to obtain perfect CSI of the link from the IRS to eavesdropper. The reasons are as follows. On one hand, the existence of channel estimation errors and quantization errors can result in imperfect CSI estimation \cite{rq2}. On the other hand, since the locations of the eavesdroppers are unknown and there is no cooperation between the legitimate user and the eavesdroppers,  perfect CSI is almost impossible to obtain \cite{rfzar1}.
Imperfect CSI can significantly deteriorate the beamforming and IRS performance. Thus, it is of crucial importance to design robust secure beamforming and phase shift matrix for IRS assisted cooperative jamming (CJ) communication networks.  

Motivated by the above-mentioned facts, in this paper, robust secure beamforming and phase shift matrix are designed for an IRS assisted MISO network with an independent cooperative jamming user. The energy efficiency maximization framework is formulated. To the author’s best knowledge, this is the first work that considers robust beamforming and cooperative jamming design in IRS-assisted MISO networks with CJ and that studies the energy efficiency maximization problems in this type of network.
\vspace{-0.15in}
\subsection{Related Work and Motivation}
Optimal beamforming design plays an important role in the improvement of secure performance in wireless communication networks by using physical layer security. The related works can be classified into three categories, namely, secure beamforming design in conventional MISO networks with CJ under perfect CSI \cite{mcj1}-\cite{mcj9}, robust secure beamforming design in conventional MISO networks with CJ under imperfect CSI \cite{rar3}-\cite{rmj7}, and secure beamforming design in IRS-assisted secure wireless networks \cite{rirs7}-\cite{rirs5}. 

For the conventional MISO network secure communication with perfect CSI,  the beamforming and jamming design were jointly optimized to achieve different objectives, e.g., the secrecy rate maximization of users \cite{mcj1}-\cite{mcj6}, the minimization of energy consumption \cite{mcj7}, and the system efficiency maximization \cite{mcj8}, \cite{mcj9}.
Specifically, in \cite{mcj1}, the authors exploited the CJ for multiple users via broadcast channels to enhance the secure performance with the help of a friendly jammer. The optimal CJ was designed to keep the achieved SINR at the eavesdroppers below the threshold to guarantee that the transmission from the base station to the legitimate users is confidential. 
To achieve a higher secrecy rate performance, in \cite{mcj4}, Park \textit{et al}. investigated a single relay assisted secure communication network. By using CJ to prevent the eavesdropper from intercepting the source message, they proposed three jamming power allocation strategies to minimize the outage probability of the secrecy rate. 
Different from the single relay system, a wireless network with multiple relays was considered in \cite{mcj5}. A two-slot cooperative relaying scheme was proposed to maximize the secrecy rate.
The access method is another key element for increasing the system secrecy rate.
The authors in \cite{mcj6} studied the secrecy rate maximization problem in an orthogonal frequency division multiple (OFDM) system with a potential eavesdropper. With the assistance of a cooperative jammer, the approaches they proposed can significantly improve the secrecy rate by jointly optimizing the transmit power and time allocation. 
While the works in \cite{mcj1}-\cite{mcj6} aim to achieve a higher secrecy rate, they only consider one performance metric therefore may not be able to achieve a good tradeoff between conflicting performance goals such as high rate and low energy consumption.
Recently, the authors in \cite{mcj7} considered secure resource allocations for OFDM networks under scenarios with and without CJ. The joint optimization problem of subcarrier assignments and power allocations subject to a limited power budget at the relay was solved to maximize the secrecy sum-rate and save energy.
Different from the works in \cite{mcj1}-\cite{mcj7}, energy-efficient secure communication was considered in \cite{mcj8}. By using two jamming strategies, namely, beamforming and cooperative diversity, they demonstrated that a cooperative diversity strategy is desirable. Significant energy efficiency can be achieved by selectively switching between the two strategies.
Besides the strategy selection, the mode switch can also improve the energy efficiency.
In \cite{mcj9}, the authors proposed an intermittent jamming strategy where a jammer alternates between jamming and non-jamming modes during the legitimate transmission. By jointly measuring security requirements and energy costs, they formulated and solved an optimization problem with respect to the jamming duration proportion and jamming power.

In practice, the perfect CSI is not always available at the transmitter. The secure network designs presented above are not suitable for imperfect CSI cases. Thus, to achieve robust design of the secure communication network, the beamforming design problems with channel estimation error have been considered \cite{rar3}-\cite{rmj7}. 
The authors in \cite{rar3} studied robust transmission schemes with a single eavesdropper for MISO networks. Both the cases of direct transmission and CJ were investigated with imperfect CSI for the eavesdropper links. Robust transmission covariance matrices were obtained by solving the worst-case secrecy rate maximization.
For the MISO system with multiple eavesdroppers, Ma \textit{et al}. in \cite{rar5} investigated a robust quality-of-service (QoS)-based and secrecy rate-based secure transmission design. By jointly optimizing the transmit beamforming vector and the covariance matrix of jamming signals under individual power constraints, they proposed an algorithm for each problem through semidefinite relaxation (SDR). 
In \cite{rmj5}, the authors aimed to minimize the total transmit power by jointly designing the beamforming vector at the transmitter and AN covariance at jammer under the reliability and secrecy constraints for all the possible distributions of CSI errors. 
Su \textit{et al}. in \cite{rmj6} proposed a novel robust beamforming strategy for the direct transmission NOMA and cooperative jamming NOMA to minimize the worst-case sum power subject to secrecy rate constraint. 
In \cite{rmj2}, Feng \textit{et al}. investigated cooperative secure beamforming for simultaneous wireless information and power transfer (SWIPT) in AF relay networks with imperfect CSI. They proposed a joint cooperative beamforming (CB) and energy signal (ES) scheme to maximize the secrecy rate under both the power constraints and the wireless power transfer constraint.
In \cite{rmj3}, Chu \textit{et al.} studied a MISO secrecy network with CJ and SWIPT to maximize the minimum harvested energy subject to the total power constraints while guaranteeing the minimum secrecy rate. By incorporating the norm-bound channel uncertainty model, they proposed a joint design of the robust secure transmission scheme which outperforms the separate AN-aided or CJ-aided schemes.
By considering the secrecy rate and consumed energy of the robust secure communication network simultaneously, the tradeoff between them can be investigated to achieve the maximum energy efficiency.
In \cite{rfzar1}, a MISO cognitive radio downlink network with SWIPT was studied. The tradeoff was elucidated between the secrecy rate and the harvesting energy under the max-min fairness criterion. 
The joint design of the beamforming vector and the artificial noise covariance matrix were investigated in \cite{rmj7} for the MISO multiple-eavesdropper SWIPT systems. The secrecy energy efficiency maximization problem was formulated and two suboptimal solutions were proposed based on the heuristic beamforming techniques.

Recently, the IRS-assisted MISO secure network has attracted increasingly elevated attention. The beamforming and phase shift matrix design schemes for different objectives were proposed in \cite{rd01}-\cite{rirs5}.
{
For the multi-user network, in \cite{rd01}, the authors investigated the symbol-level precoding in IRS-assisted multiuser MISO systems to minimize the transmit power while guaranteeing the information transmissions. In \cite{rd02}, the authors considered the downlink multigroup multicast communication systems assisted by an IRS. By optimizing the precoding matrix and the reflection coefficients, the sum rate of all the multicasting groups was maximized. 
For the multi-IRS deployment problems, in \cite{rd11}, the deploying strategies for IRS were investigated for a single-cell multiuser system aided by multiple IRSs. It was shown that the IRS-aided system outperforms the full-duplex relay-aided counterpart system and that the deployment strategies and the elements of IRS have significant influence on the achievable spatial throughput.
In \cite{rd12}, the authors analyzed the impact of the deployment of IRS on the downlink throughput and showed that IRS density can significantly enhance the signal power at the expense of only a marginally increasing interference.
}
To investigate the secrecy rate gain brought by IRS, in \cite{rirs7}, Yu \textit{et al.} considered an IRS-assisted secure MISO wireless system. To maximize the secrecy rate, both the beamformer and the IRS phase shift matrix were jointly optimized based on the block coordinate descent (BCD) and minimization maximization techniques. 
By combining the AN technique, in \cite{rirs6}, Xu \textit{et al}. studied resource allocation design to maximize the system sum secrecy rate. By jointly optimized the phase shift matrix, the beamforming vectors, and the AN covariance matrix, the authors developed an efficient suboptimal algorithm based on alternating optimization, successive convex approximation, SDR, and manifold optimization.
In \cite{rirs2}, by jointly optimizing the beamformers at the BS and reflecting coefficients at the IRS, the authors formulated a minimum-secrecy-rate maximization problem under various practical constraints that captured the scenarios of both continuous and discrete reflecting coefficients of the reflecting elements. 
Since IRS can not only help increase the secrecy rate but also save more energy for the network, the joint optimization of rate and power was also studied.
By considering the power consumption, in \cite{rirs3}, the authors focused on maximizing the system secrecy rate subject to the source transmission power constraint and the unit modulus constraints imposed on phase shifts at the IRS. 
Furthermore, in \cite{rirs1}, the authors proposed a power-efficient scheme to optimize the secure transmit power allocation and the surface reflecting phase shift to minimize the transmit power subject to the secrecy rate constraint.
In \cite{rirs4}, the authors proposed different methods to minimize the system's energy consumption in cases of rank-one and full-rank access point (AP)-IRS links. 
In \cite{rirs5}, secure wireless information and power transfer with the IRS was proposed for a MISO system. Under the secrecy rate and the reflecting phase shifts of IRS constraints, the secure transmit beamforming at the access point and phase shifts at IRS were jointly optimized to maximize the harvested power of the energy harvesting receiver. 

Although beamforming design problems in CJ assisted secure MISO networks \cite{mcj1}-\cite{mcj9}, IRS-enabled secure communication systems \cite{rirs7}-\cite{rirs5} under perfect CSI, and robust beamforming design problems in CJ assisted secure MISO networks under the imperfect CSI \cite{rar3}-\cite{rmj7} have been investigated, few studies have been conducted for beamforming, friendly jamming and phase shift matrix design in IRS assisted wireless MISO networks. Moreover, energy efficiency optimization based on perfect CSI proposed in the above-mentioned works are not applicable to the imperfect CSI since the channel estimated errors can have a big impact on the performance of both base station and friendly jammer. Furthermore, with imperfect CSI, the application of IRS into the MISO network with friendly jamming can face more challenges that have not been considered in the works mentioned above. Thus, in order to improve energy efficiency performance and achieve robustness against the uncertainty introduced by the imperfect CSI, it is of crucial importance to study robust beamforming, friendly jamming, and phase shift matrix design problems in IRS-aided MISO networks. These problems are normally challenging to tackle due to two reasons. There exists dependence among different variables that makes the problems non-convex. The imperfect CSI model further increases the complexity of the problems by introducing the uncertainty constraints to the optimization problems.
\vspace{-0.15in}
\subsection{Contribution and Organization}
Our preliminary work in \cite{irsqun} only  considered the energy efficiency maximization problem under the perfect CSI model. Motivated by the above-mentioned observations, in this paper, the energy efficiency maximization problems are studied in an IRS-assisted MISO network with cooperative jamming under both perfect and imperfect CSI models. The corresponding  robust design  to address channel uncertainty is also provided.  The major contributions of this paper are summarized as follows.
\subitem 1) We investigate the joint design of information beamforming, cooperative jamming, and phase shift matrix to maximize the energy efficiency in an IRS-assisted secure network with eavesdroppers under the perfect CSI model. The problem is challenging to solve due to its non-convexity and coupling of the beamforming vector with the IRS phase shift matrix. An alternating optimization algorithm is proposed to solve the non-convex fractional problem by using SDR.
\subitem 2) For the IRS aided MISO network under imperfect CSI model, the estimated channel error results in the uncertainty to the system and brings more difficulties for beamforming and phase shift matrix design compared with the perfect CSI case.  To deal with this uncertainty, the bounded channel error model is considered and the $\mathcal{S}$-procedure method is applied for optimizing the robust beamforming and IRS phase shift matrix to maximize the energy efficiency.  
\subitem 3) The simulation results show that the proposed method with the perfect CSI can achieve the highest energy efficiency among all the benchmark methods.
Moreover, it is found that there is a tradeoff between secrecy rate and the consumed energy. Furthermore, it is shown that the exploitation of IRS is beneficial for improving energy efficiency even under the imperfect CSI case. 

The rest of the paper is organized as follows. In Section II,  the system model is provided.  In Section III, the energy efficiency maximization problem under perfect CSI and its solution are presented. In Section IV, the energy efficiency maximization problem under the imperfect CSI case is studied. Simulation results are given in Section V. Finally, the paper is concluded in Section VI. 

\textit{Notation:} $\mathbb{C}^{M\times N}$ denotes the $M \times N $ complex-valued matrices space. $\mathcal{CN}(\mu, \sigma^2)$ denotes the distribution of complex Gaussian random variable with mean $\mu$ and variance $\sigma^2$. For a square matrix $\mathbf{X}$, the trace of $\mathbf{X}$ is denoted as $\text{Tr}(\mathbf{X})$ and rank($\mathbf{X}$) denotes the rank of matrix $\mathbf{X}$. $\angle(x)$ denotes the phase of complex number $x$. Matrices and vectors are denoted by boldface capital letters and boldface lower case letters. $[x]^+$ denotes the maximum between $0$ and $x$.

\vspace{-0.2in}
\section{System Model}

As shown in Fig. 1, an IRS assisted wireless communication system is considered. A multi-antenna base station transmits the confidential information to a single-antenna legitimate user. At the same time, $K$ eavesdroppers (Eves) are trying to intercept the information from the base station. In order to improve the security, a friendly jammer intentionally issues the jamming signals. It is assumed that both the base station and the jammer are equipped with $N$ antennas, and the IRS has $M$ reflecting elements. Each Eve is equipped with a single antenna. 
	\begin{figure}[h]
			\setlength{\abovedisplayskip}{3pt}
		\setlength{\belowdisplayskip}{3pt}
	\centering
	\includegraphics[width=3.0in]{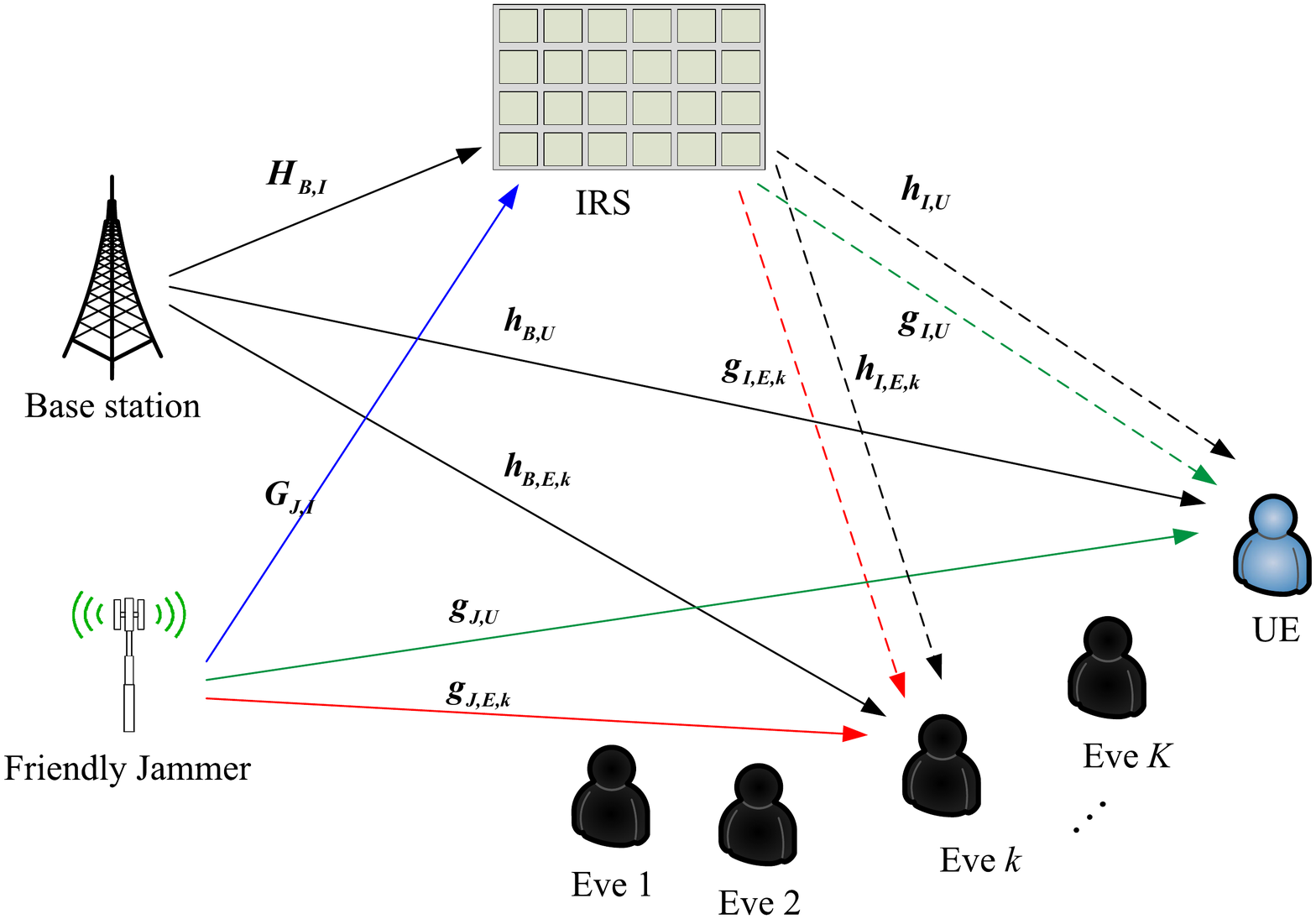}	\label{model}
	\caption{An IRS-aided MISO wireless network with a friendly jammer.}
\end{figure}

The base band equivalent channel from the base station to the IRS, base station to the user, and base station to the $k$th Eve are denoted as $\mathbf{H}_{B,I} \in \mathbb{C}^{M \times N}$, $\mathbf{h}_{B,U} \in \mathbb{C}^{1 \times N}$, and $\mathbf{h}_{B,E,k} \in \mathbb{C}^{1 \times N}$, respectively. The baseband equivalent channel from the Jammer to the IRS, Jammer to user,  and Jammer to Eve $k$ are denoted as $\mathbf{G}_{J,I} \in \mathbb{C}^{M \times N}$, $\mathbf{g}_{J,U} \in \mathbb{C}^{1 \times N}$, and $\mathbf{g}_{J,E,k} \in \mathbb{C}^{1 \times N}$, respectively. The channel from the IRS to the user and Eve $k$ are denoted as $\mathbf{h}_{I,U}$, $\mathbf{h}_{I,E,k}$, $\mathbf{g}_{I,U}$, and $\mathbf{g}_{I,E,k}$, respectively, each of which is  a $1 \times M$ complex vector. 
{
The performance achieved under the perfect CSI case can serve as an upper bound for the proposed secure communication design.
According to the works in \cite{wqqirs}, \cite{rd16}, the Eves may be legitimate users in the past but cannot access the confidential information in the current communication process or the base station does not want to send confidential information to those users.  To guarantee the communication security, the system has to treat those receivers as potential eavesdroppers. Therefore, the perfect CSI of those Eves can be acquired.}

It is assumed that the channel information between the IRS and the user is available at both the base station and the jammer. However, since eavesdroppers normally try to hide their existence from the base station, it is difficult to obtain the perfect CSI between Eves and base station. In practice, the CSI knowledge of the links from the IRS to Eves is not accurate. This can also be caused by channel estimation and quantization errors. In order to develop a robust scheme under the imperfect CSI case, the worst case channel uncertainty model is considered. The bounded CSI error models for the channel vector $\mathbf{h}_{I,E,k}$, and $\mathbf{g}_{I,E,k}$ are given as
\begin{subequations}
			\setlength{\abovedisplayskip}{3pt}
	\setlength{\belowdisplayskip}{3pt}
	\begin{alignat}{5}
	\mathbf{h}_{I,E,k}=&\overline{\mathbf{h}}_{I,E,k}+\Delta\mathbf{h}_{I,E,k},~\mathcal{H}_{I,E,k}\\ \nonumber
	=&\{\Delta\mathbf{h}_{I,E,k}\in\mathbb{C}^{M\times 1}:\Delta\mathbf{h}_{I,E,k}^H\Delta\mathbf{h}_{I,E,k}\le\xi_{I,E,k}^2 \},\\
	\mathbf{g}_{I,E,k}=&\overline{\mathbf{g}}_{I,E,k}+\Delta\mathbf{g}_{I,E,k},~\mathcal{G}_{I,E,k}\\\nonumber
	=&\{\Delta\mathbf{g}_{I,E,k}\in\mathbb{C}^{M\times 1}: \Delta\mathbf{g}_{I,E,k}^H\Delta\mathbf{g}_{I,E,k}\le\xi_{J,E,k}^2 \},
	\end{alignat}
\end{subequations}
where $\overline{\mathbf{h}}_{I,E,k}$, and $\overline{\mathbf{g}}_{I,E,k}$ are the estimated values  of the channel vectors $\mathbf{h}_{I,E,k}$, and $\mathbf{g}_{I,E,k}$, respectively. $\mathcal{H}_{I,E,k}$, and $\mathcal{G}_{I,E,k}$ denote the uncertainty regions of $\mathbf{h}_{I,E,k}$, and $\mathbf{g}_{I,E,k}$, respectively. $\Delta\mathbf{h}_{I,E,k}$, and $\Delta\mathbf{g}_{I,E,k}$ represent the channel estimation errors. $\xi_{I,E,k}$, and $\xi_{J,E,k}$ are the radius of the uncertainty region $\mathcal{H}_{I,E,k}$, and $\mathcal{G}_{I,E,k}$, respectively \cite{rfzar1}. 

In this paper, the IRS adjusts its elements to maximize the combined incident signal for  the legitimate user. The diagonal phase-shift matrix can be denoted as $\mathbf{\Theta} = \text{diag}(\exp({j\theta_{1}}), \exp({j\theta_{2}}), $ $\cdots, \exp({j\theta_{M}}))$, wherein its main diagonal, $\theta_{m}\in [0,2\pi)$, denotes the phase shift on the combined incident signal by its $m$th element, $m=1,2,...,M$ \cite{cwh3}. 

The transmitted signal from the base station to the user is given as $\mathbf{x}_B = \mathbf{f}_1 s_1$ and the jamming signal from the jammer is given as $\mathbf{x}_J =\mathbf{f}_2 s_2$, where
$s_1  \sim \mathcal{CN}(0,1)$ and $s_2 \sim \mathcal{CN}(0,1)$ denote the independent information and jamming signal, respectively. 	$\mathbf{f}_1 \in \mathbb{C}^{N \times 1}$ and $\mathbf{f}_2 \in \mathbb{C}^{N \times 1}$ denote the beamforming and jamming precode vectors, respectively. Let $P_{1,max}$ and $P_{2,max}$ denote the maximum transmit power available at base station and jammer. We have $(\mathbf{f}_1^{H}\mathbf{f}_1) \le P_{1,max}$ and $(\mathbf{f}_2^{H}\mathbf{f}_2) \le P_{2,max}$.
The signal received at legitimate user and Eve $k$ can be  respectively given as 
\be
		\setlength{\abovedisplayskip}{3pt}
\setlength{\belowdisplayskip}{3pt}
\begin{aligned}
\mathbf{y}_{U}&=(\mathbf{h}_{B,U}^{H}+\mathbf{h}_{I,U}^{H}\mathbf{\Theta}\mathbf{H}_{B,I})\mathbf{f}_1 s_1\\
&+(\mathbf{g}_{J,U}^{H}+\mathbf{g}_{I,U}^{H}\mathbf{\Theta}\mathbf{G}_{J,I})\mathbf{f}_2 s_2+n_U,
\end{aligned}
\ee
and
\be
		\setlength{\abovedisplayskip}{3pt}
\setlength{\belowdisplayskip}{3pt}
\begin{aligned}
\mathbf{y}_{E,k}&=(\mathbf{h}_{B,E,k}^{H}+\mathbf{h}_{I,E,k}^{H}\mathbf{\Theta}\mathbf{H}_{B,I})\mathbf{f}_1 s_1\\
&+(\mathbf{g}_{J,E,k}^{H}+\mathbf{g}_{I,E,k}^{H}\mathbf{\Theta}\mathbf{G}_{J,I})\mathbf{f}_2 s_2+n_{E,k},
\end{aligned}
\ee
where $n_U$ and $n_{E,k} \sim \mathcal{CN}(0,\sigma^2)$ are the complex additive white Gaussian noise (AWGN). 
Thus, the signal of interference plus noise ratio (SINR) of the legitimate user and Eve $k$ can be given as 
\be
		\setlength{\abovedisplayskip}{3pt}
\setlength{\belowdisplayskip}{3pt}
\gamma_U = \frac{|(\mathbf{h}_{B,U}^{H}+\mathbf{h}_{I,U}^{H}\mathbf{\Theta}\mathbf{H}_{B,I})\mathbf{f}_1|^2}{|(\mathbf{g}_{J,U}^{H}+\mathbf{g}_{I,U}^{H}\mathbf{\Theta}\mathbf{G}_{J,I})\mathbf{f}_2|^2+\sigma^2},
\ee
and
\be
		\setlength{\abovedisplayskip}{3pt}
\setlength{\belowdisplayskip}{3pt}
\gamma_{E,k} = \frac{|(\mathbf{h}_{B,E,k}^{H}+\mathbf{h}_{I,E,k}^{H}\mathbf{\Theta}\mathbf{H}_{B,I})\mathbf{f}_1|^2}{|(\mathbf{g}_{J,E,k}^{H}+\mathbf{g}_{I,E,k}^{H}\mathbf{\Theta}\mathbf{G}_{J,I})\mathbf{f}_2|^2+\sigma^2}.
\ee

The  achievable secrecy rate is defined as 
\be
		\setlength{\abovedisplayskip}{3pt}
\setlength{\belowdisplayskip}{3pt}
R_S=[R_U-R_E]^+=[B\log_2(1+\gamma_U)-\max_{k \in K}B\log_2(1+\gamma_{E,k})]^+.
\ee

The energy consumed by the base station and the jammer consists of the transmit power and the circuit power consumption $P_{BS}$ and $P_{G}$. The power consumed by the IRS is denoted as $P_{IRS}$. Thus, the total power consumed in the system can be given as
\be
		\setlength{\abovedisplayskip}{3pt}
\setlength{\belowdisplayskip}{3pt}
P_{tot}= \zeta (\mathbf{f}_1^{H}\mathbf{f}_1+\mathbf{f}_2^{H}\mathbf{f}_2)+P_{BS}+P_{G}+P_{IRS},
\ee
where $\zeta$ is the amplifier coefficient.

According to \cite{cwh3}, the energy efficiency is defined as
\be
		\setlength{\abovedisplayskip}{3pt}
\setlength{\belowdisplayskip}{3pt}
\eta =\frac{[B\log_2(1+\gamma_U)-\max_{k \in K}B\log_2(1+\gamma_{E,k})]^+}{ \zeta (\mathbf{f}_1^{H}\mathbf{f}_1+\mathbf{f}_2^{H}\mathbf{f}_2)+P_{BS}+P_{G}+P_{IRS}}.
\ee

In order to maximize the energy efficiency, the beamforming and jamming vectors and the phase shift matrix are jointly optimized. Since the energy efficiency maximization problem is extremely challenging under the imperfect CSI case, the problem is firstly studied under the perfect CSI case in order to provide some meaningful insights in Section \ref{perfect}. Based on the results obtained in Section \ref{perfect}, the energy efficiency maximization problem is further studied under the imperfect CSI in Section \ref{imperfect}.  
\section{System Design With Perfect CSI}
\label{perfect}
In this section, the energy efficiency maximization problem with perfect CSI is studied by jointly optimizing the beamforming vector, jamming vector, and phase shift matrix. An alternating algorithm is proposed to tackle the challenging non-convex problem.
\vspace{-0.2in}
\subsection{Problem Formulation}
Under the perfect CSI model, the energy efficiency maximization problem is formulated as 
\begin{subequations}
		\setlength{\abovedisplayskip}{3pt}
	\setlength{\belowdisplayskip}{3pt}
	\label{P0}
	\begin{alignat}{5}
	\text{P}{_1:}~ &\max_{\mathbf{f}_1,\mathbf{f}_2,\mathbf{ \Theta}}~\eta\nonumber\\
	s.t.~~&\mathbf{f}_1^{H}\mathbf{f}_1 \le P_{1,max},~\mathbf{f}_2^{H}\mathbf{f}_2 \le P_{2,max},\\
	&R_s \ge R_{th},\\
	&|\exp({j\theta_{m}})|=1,
	\end{alignat}
\end{subequations} 
where $R_{th}$ is the minimum required secure rate threshold. It is evident that problem $\text{P}_1$ is non-convex due to the fractional structure of the objective function and the non-convex constraints. In order to tackle it, an alternating algorithm is proposed to solve this problem.


The problem $\text{P}_{1}$ is non-convex due to the coupling of the beamforming vector, jamming vector and IRS phase shift matrix. 
By introducing ${\mathbf{w}^H}=[w_1, w_2, \cdots, w_M]$,  one has $\mathbf{h}_{I,j}^{H}\mathbf{\Theta}\mathbf{H}_{B,I}={\mathbf{w}^H}\mathbf{H}{_{I,j}}$, where $w_m=\exp({j\theta_{m}})$, $\mathbf{H}_{I,j}=\text{diag}(\mathbf{h}_{I,j}^H)\mathbf{H}_{B,I},~j\in \{U,(E,k)\}$. The interference from the jammer can be denoted as $\mathbf{g}_{I,j}^{H}\mathbf{\Theta}\mathbf{G}_{J,I}={\mathbf{w}^H}\mathbf{G}_{I,j}$, where $\mathbf{G}{_{I,j}}=\text{diag}(\mathbf{g}_{I,j}^H)\mathbf{G}_{J,I},~j\in \{U,(E,k)\}$. { Thus, the SINRs of user and Eve $k$ are given as $\gamma_j=\frac{a_0|\overline{\mathbf{w}}^H\mathbf{H}_{j}\mathbf{f}_1|^2}{a_0|\overline{\mathbf{w}}^H\mathbf{G}_{j}\mathbf{f}_2|^2+1},~~j\in\{U,(E,k)\}$, where $a_0=1/\sigma^2$, $\mathbf{H}_{j}=\left[ {\begin{array}{*{20}{c}}
	{{\mathbf{H}_{I,j}}}  \\
	{{\mathbf{h}_{B,j}}}  \\
	\end{array}} \right]$, $\mathbf{G}_{j}=\left[ {\begin{array}{*{20}{c}}
	{{\mathbf{G}_{I,j}}}  \\
	{{\mathbf{g}_{J,j}}}  \\
	\end{array}} \right]$, $\overline{\mathbf{w}}^H=\exp({j\overline{w}})[\mathbf{w}^H,1]$ and $\overline{w}$ is an arbitrary phase rotation.
The problem can be transformed into 
\begin{subequations}
			\setlength{\abovedisplayskip}{3pt}
	\setlength{\belowdisplayskip}{3pt}
	\begin{alignat}{5}
		\setlength{\abovedisplayskip}{3pt}
	\setlength{\belowdisplayskip}{3pt}
	\begin{split}
	\text{P}{_{1.1}:}~ &\max_{\mathbf{f}_1,\mathbf{f}_2,\overline{\mathbf{w}}}~\frac{1}{P_{tot}}\{\frac{B}{\ln2}\ln(1+\frac{a_0|\overline{\mathbf{w}}^H\mathbf{H}_{U}\mathbf{f}_1|^2}{a_0|\overline{\mathbf{w}}^H\mathbf{G}_{U}\mathbf{f}_2|^2+1})\\
	&-\max_{k \in K}\frac{B}{\ln2}\ln(1+\frac{a_0|\overline{\mathbf{w}}^H\mathbf{H}_{E,k}\mathbf{f}_1|^2}{a_0|\overline{\mathbf{w}}^H\mathbf{G}_{E,k}\mathbf{f}_2|^2+1})\}
	\end{split}\nonumber\\
	s.t.~~& (9\text{a}), (9\text{c}),\nonumber\\\
	\begin{split}
	&\frac{B}{\ln2}\ln(1+\frac{a_0|\overline{\mathbf{w}}^H\mathbf{H}_{U}\mathbf{f}_1|^2}{a_0|\overline{\mathbf{w}}^H\mathbf{G}_{U}\mathbf{f}_2|^2+1})\\
	&-\max_{k \in K}\frac{B}{\ln2}\ln(1+\frac{a_0|\overline{\mathbf{w}}^H\mathbf{H}_{E,k}\mathbf{f}_1|^2}{a_0|\overline{\mathbf{w}}^H\mathbf{G}_{E,k}\mathbf{f}_2|^2+1}) \ge R_{th}.
	\end{split}
	\end{alignat}
\end{subequations}}
The problem $\text{P}_{1.1}$ is yet still non-convex. In order to tackle it, the beamforming and jamming vectors can be optimized for a given $\overline{\mathbf{w}}$, and then $\overline{\mathbf{w}}$ can be optimized for the obtained optimal $\mathbf{f}_1$ and $\mathbf{f}_2$. This process iteratively continues till convergence. 
\vspace{-0.3cm}
\subsection{Optimizing the Beamforming for a Given $\overline{\mathbf{w}}$}
In this section, we solve the problem $\text{P}_{1.1}$ to achieve the optimal secure transmit beamformer $\mathbf{f}_1$ and jammer vector $\mathbf{f}_2$ for a given $\overline{\mathbf{w}}$.
Let $\mathbf{\overline{h}}_{U}^H=\mathbf{\overline{w}}^H\mathbf{H}_{U}$, $\mathbf{\overline{g}}_{U}^H=\mathbf{\overline{w}}^H\mathbf{G}_{U}$,  $\mathbf{\overline{h}}_{E,k}^H=\mathbf{\overline{w}}^H\mathbf{H}_{E,k}$, and $\mathbf{\overline{g}}_{E,k}^H=\mathbf{\overline{w}}^H\mathbf{G}_{E,k}$. {The problem $\text{P}_{1.1}$ can be transformed into 
\begin{subequations}
			\setlength{\abovedisplayskip}{3pt}
	\setlength{\belowdisplayskip}{3pt}
	\begin{alignat}{5}
		\setlength{\abovedisplayskip}{3pt}
	\setlength{\belowdisplayskip}{3pt}
	\begin{split}
	\text{P}{_{1.2}:}~ &\max_{\mathbf{f}_1,\mathbf{f}_2}~\frac{1}{P_{tot}}\{\frac{B}{{\ln 2}}\ln(1 + \frac{{{a_0}|\overline {\mathbf{h}} _{U}^H{{\mathbf{f}}_1}{|^2}}}{{{a_0}|\overline {\mathbf{g}} _{U}^H{{\mathbf{f}}_2}{|^2} + 1}})\\
		&- {\max _{k \in K}}\frac{B}{{\ln 2}}\ln(1 + \frac{{{a_0}|\overline {\mathbf{h}} _{E,k}^H{{\mathbf{f}}_1}{|^2}}}{{{a_0}|\overline {\mathbf{g}} _{E,k}^H{{\mathbf{f}}_2}{|^2} + 1}})\}
	\end{split}\nonumber\\
	s.t.~~
	\begin{split}
	&\frac{B}{{\ln 2}}\ln(1 + \frac{{{a_0}|\overline {\mathbf{h}} _{U}^H{{\mathbf{f}}_1}{|^2}}}{{{a_0}|\overline {\mathbf{g}} _{U}^H{{\mathbf{f}}_2}{|^2} + 1}})\\
	&- {\max _{k \in K}}\frac{B}{{\ln 2}}\ln(1 + \frac{{{a_0}|\overline {\mathbf{h}} _{E,k}^H{{\mathbf{f}}_1}{|^2}}}{{{a_0}|\overline {\mathbf{g}} _{E,k}^H{{\mathbf{f}}_2}{|^2} + 1}}) \ge R_{th},
	\end{split}\\
	&\mathbf{f}_1^{H}\mathbf{f}_1 \le P_{1,max},\mathbf{f}_2^{H}\mathbf{f}_2 \le P_{2,max}.
	\end{alignat}
\end{subequations}}

Let $|\overline{\mathbf{h}}_{j}^H\mathbf{f}_1|^2 = \rm{Tr}(\overline{\mathbf{H}}_{j}\mathbf{f}_1\mathbf{f}_1^H)$ and $|\overline {\mathbf{g}} _{j}^H\mathbf{f}_2|^2 = \rm{Tr}(\overline{\mathbf{G}}_{j}\mathbf{f}_2\mathbf{f}_2^H)$. By defining $\overline{\mathbf{H}}_{j}=\mathbf{\overline{h}}_{j}\mathbf{\overline{h}}_{j}^H$, $\overline{\mathbf{G}}_{j}=\mathbf{\overline{g}}_{j}\mathbf{\overline{g}}_{j}^H$, $j\in\{U,(E,k)\}$, $\mathbf{F}_1 = \mathbf{f}_1\mathbf{f}_1^H$ and $\mathbf{F}_2 = \mathbf{f}_2\mathbf{f}_2^H$, one has $\mathbf{F}_1\succeq0$, $\mathbf{F}_2\succeq0$ and $\text{rank}(\mathbf{F}_1)=\text{rank}(\mathbf{F}_2)=1$. 
The rank-$1$ constraint makes  the problem difficult to solve. Thus, we apply the SDR method to relax the constraints. The problem $\text{P}_{1.2}$  is thus expressed as 
\begin{subequations}
	\begin{alignat}{5}
		\setlength{\abovedisplayskip}{3pt}
	\setlength{\belowdisplayskip}{3pt}
	\begin{split}
	&\text{P}{_{1.3}:}~\max_{\mathbf{F}_1,\mathbf{F}_2}~\frac{1}{P_{tot}}(\frac{B}{{\ln 2}}\ln(1 + \frac{a_0\rm{Tr}(\overline{\mathbf{H}} _{U}\mathbf{F}_1)}{a_0\rm{Tr}(\overline{\mathbf{G}}_{U}\mathbf{F}_2) + 1})\\
		&- {\max_{k \in K}}\frac{B}{{\ln 2}}\ln(1 + \frac{a_0\rm{Tr}(\overline{\mathbf{H}}_{E,k}\mathbf{F}_1)}{a_0\rm{Tr}(\overline{\mathbf{G}}_{E,k}\mathbf{F}_2) + 1}))
	\end{split}\nonumber\\
	s.t.~~&(\mathbf{F}_1,\mathbf{F_2})\in \mathcal{F},\\
	\begin{split}
	&\frac{B}{{\ln 2}}\ln(1 + \frac{a_0\rm{Tr}(\overline{\mathbf{H}} _{U}\mathbf{F}_1)}{a_0\rm{Tr}(\overline{\mathbf{G}}_{U}\mathbf{F}_2) + 1})\\
	&- {\max _{k \in K}}\frac{B}{{\ln 2}}\ln(1 + \frac{a_0\rm{Tr}(\overline{\mathbf{H}}_{E,k}\mathbf{F}_1)}{a_0\rm{Tr}(\overline{\mathbf{G}}_{E,k}\mathbf{F}_2) + 1}) \ge R_{th},
	\end{split}
	\end{alignat}
\end{subequations}
where $\mathcal{F}=\{(\mathbf{F}_1,\mathbf{F}_2)|\text{Tr}(\mathbf{F}_1)\le P_{1,\max},~\text{Tr}(\mathbf{F}_2)\le P_{2,\max},~\mathbf{F}_1 \succeq0,~\mathbf{F}_2 \succeq0\}$. However, the problem $\text{P}_{1.3}$ is still a non-convex problem due to the objective function and the non-convex second constraint with respect to $\mathbf{F}_1$ and $\mathbf{F}_2$. To solve this, the following lemma is applied \cite{lemma1}.

$\mathbf{Lemma~1}$: By introducing the function $\phi(t)=-tx+\ln t+1$ for any $x > 0$, one has
\be
	\setlength{\abovedisplayskip}{3pt}
\setlength{\belowdisplayskip}{3pt}
-\ln x=\max_{t>0}\phi(t).
\ee
The optimal solution can be achieved at $t=1/x$. The upper bound can be given by using Lemma $1$ as $\phi(t)$. By setting $x=a_0\text{Tr}(\overline{\mathbf{G}}_{U}\mathbf{F}_2)+1$, and $t=t_U$, one has 
\be
	\setlength{\abovedisplayskip}{3pt}
\setlength{\belowdisplayskip}{3pt}
\begin{split}
	R_U\frac{\ln2}{B}&=[{\ln(a_0\text{Tr}(\overline{\mathbf{H}}_{U}\mathbf{F}_1)+a_0\text{Tr}(\overline{\mathbf{G}}_{U}\mathbf{F}_2)+1)}\\
	&-\ln(a_0\text{Tr}(\overline{\mathbf{G}}_{U}\mathbf{F}_2)+1)]=\max_{t_U>0}\phi_u({\mathbf{F}_1,\mathbf{F}_2, }t_U),
\end{split}
\ee
where $\phi_U({\mathbf{F}_1,\mathbf{F}_2, }t_U)={\ln(a_0\text{Tr}(\overline{\mathbf{H}}_{U}\mathbf{F}_1)+a_0\text{Tr}(\overline{\mathbf{G}}_{U}\mathbf{F}_2)+1)}$ $-t_U(a_0\text{Tr}(\overline{\mathbf{G}}_{U}\mathbf{F}_2)+1)+\ln t_U +1$.

In the same way, let $x=a_0\text{Tr}(\overline{\mathbf{H}}_{E,k}\mathbf{F}_1)+a_0\text{Tr}(\overline{\mathbf{G}}_{E,k}\mathbf{F}_2)+1$ and $t=t_{E,k}$, one has
\be
	\setlength{\abovedisplayskip}{3pt}
\setlength{\belowdisplayskip}{3pt}
\begin{split}
	R_{E,k}\frac{\ln2}{B}&=[\ln(a_0\text{Tr}(\overline{\mathbf{H}}_{E,k}\mathbf{F}_1)+a_0\text{Tr}(\overline{\mathbf{G}}_{E,k}\mathbf{F}_2)+1)\\
	&-\ln(a_0\text{Tr}(\overline{\mathbf{G}}_{E,k}\mathbf{F}_2)+1)]\\
	&=\min_{t_{E,k}>0}\phi_{E,k}({\mathbf{F}_1,\mathbf{F}_2, }t_{E,k}),
\end{split}
\ee
where $\phi_{E,k}({\mathbf{F}_1,\mathbf{F}_2, }t_{E,k})=t_{E,k}(a_0\text{Tr}(\overline{\mathbf{H}}_{E,k}\mathbf{F}_1)+a_0\text{Tr}(\overline{\mathbf{G}}_{E,k}\mathbf{F}_2)+1)-\ln(a_0\text{Tr}(\overline{\mathbf{G}}_{E,k}\mathbf{F}_2)+1)-\ln t_{E,k} -1$.
{By using Sion's minimax theorem \cite{sion}, the problem given by eq. (16) can be transformed into
\begin{subequations}
		\setlength{\abovedisplayskip}{3pt}
	\setlength{\belowdisplayskip}{3pt}
	\begin{alignat}{5}
	\begin{split}
	\text{P}_{1.4}&~~\max_{\mathbf{F}_1, \mathbf{F}_2, t_U, t_{E,k}}\frac{\phi_U({\mathbf{F}_1,\mathbf{F}_2, }t_U)-\max_{k}\phi_{E,k}({\mathbf{F}_1,\mathbf{F}_2 }t_{E,k})}{\frac{\ln2}{B}(\text{Tr}(\mathbf{F}_1+\mathbf{F}_2)+P_{BS}+P_{G}+P_{IRS})}
	\end{split}\nonumber\\
	\text{s.t.}&~~(\mathbf{F}_1, \mathbf{F}_2) \in \mathcal{F},\\
	&\phi_U({\mathbf{F}_1,\mathbf{F}_2, }t_U)-\max_{k}\phi_{E,k}({\mathbf{F}_1,\mathbf{F}_2, }t_{E,k}) \ge R_{th}\frac{\ln2}{B},\\
	&t_U, t_{E,k} \ge 0.
	\end{alignat}
\end{subequations}}
{
According to Lemma $1$, the optimal values of $t_U$ and $t_{E,k}$ can be achieved when $t_U^*=(a_0\text{Tr}(\overline{\mathbf{G}}_{U}\mathbf{F}_2)+1)^{-1}$ and $ t_{E,k}^*=(a_0\text{Tr}(\overline{\mathbf{H}}_{E,k}\mathbf{F}_1)+a_0\text{Tr}(\overline{\mathbf{G}}_{E,k}\mathbf{F}_2)+1)^{-1}$.
Here, a slack variable $l \ge \max_{k \in K} \phi_{E,k}$ is introduced. Thus, the optimization problem $\text{P}_{1.4}$ for $\mathbf{F}_1$ and $\mathbf{F}_2$ based on $t_U^*$ and $t_{E,k}^*$ can be given as
\begin{subequations}
		\setlength{\abovedisplayskip}{3pt}
	\setlength{\belowdisplayskip}{3pt}
	\begin{alignat}{5}
	\begin{split}
	\text{P}_{1.5}&~~\max_{\mathbf{F}_1, \mathbf{F}_2}\frac{\phi_U({\mathbf{F}_1,\mathbf{F}_2, }t_U^*)-l} {\frac{\ln2}{B}(\zeta\text{Tr}(\mathbf{F}_1+\mathbf{F}_2)+P_{BS}+P_{G}+P_{IRS})}\nonumber\\
	\end{split}\\
	\text{s.t.}&~~(\mathbf{F}_1, \mathbf{F}_2) \in \mathcal{F},\\
	&\phi_U({\mathbf{F}_1,\mathbf{F}_2, }t_U^*)-l \ge R_{th}\frac{\ln2}{B},\\
	&\phi_{E,k}({\mathbf{F}_1,\mathbf{F}_2, }t_{E,k}^*) \le l.
	\end{alignat}
\end{subequations}}
{The objective function of $\text{P}_{1.5}$ is now a concave function over a convex function, and the constraints are all convex, since $\phi_U({\mathbf{F}_1,\mathbf{F}_2, }t_U^*)$ is concave and $\phi_{E,k}(\mathbf{F}_1,\mathbf{F}_2,t_{E,k}^*)$ is convex. 
	It is a single ratio maximization problem and can be solved with the Dinkelbach's method \cite{rea1}\cite{rea2}. Using the Dinkelbach's method \cite{dink1}, $\text{P}{_{1.5}}$ can be solved by iteratively solving the following problem, given as
\begin{subequations}
		\setlength{\abovedisplayskip}{3pt}
	\setlength{\belowdisplayskip}{3pt}
	\begin{alignat}{5}
	\begin{split}
	\text{P}_{1.6}~~\max_{\mathbf{F}_1, \mathbf{F}_2}&\phi_U({\mathbf{F}_1,\mathbf{F}_2, }t_U^*)-l -\frac{\ln2}{B}\eta_1^* (\zeta\text{Tr}(\mathbf{F}_1+\mathbf{F}_2)\\
	&+P_{BS}+P_{G}+P_{IRS})\\
	\end{split}\nonumber\\
	&\text{s.t.}~~(17\text{a}), (17\text{b}),(17\text{c}), \nonumber
	\end{alignat}
\end{subequations}
where $\eta_1^*$ is a non-negative parameter.}
$\text{P}_{1.6}$ is convex and can be solved by using a standard convex optimization tool \cite{wqqirs}. 

After the $\mathbf{F_1}$ and $\mathbf{F_2}$ are obtained, if $\text{rank}(\mathbf{F_1})=\text{rank}(\mathbf{F_2})=1$,  $\mathbf{f_1}$ and $\mathbf{f_2}$ can be obtained from $\mathbf{F_1}=\mathbf{f_1}\mathbf{f_1^H}$ and $\mathbf{F_2}=\mathbf{f_2}\mathbf{f_2^H}$ by performing the eigenvalue decomposition. Otherwise, the Gaussian randomization can be used for recovering the approximate $\mathbf{f_1}$ and $\mathbf{f_2}$ \cite{wqqirs}. Thus, the problem $\text{P}_{1.2}$ can be solved by alternately updating $(t_U,t_{E,k})$ and $(\mathbf{f_1},\mathbf{f_2})$, which is summarized in Algorithm 1. 
\vspace{-0.15in}
\subsection{Optimizing $\mathbf{w}$ with $(\mathbf{f}_1,\mathbf{f}_2)$}
After obtaining the beamforming vectors $\mathbf{f}_1$ and $\mathbf{f}_2$, by setting $\mathbf{h}_{W,U}=\mathbf{H}_{U}\mathbf{f_1}$, $\mathbf{g}_{W,U}=\mathbf{G}_{U}\mathbf{f_2}$, $\mathbf{h}_{W,E,k}=\mathbf{H}_{E,k}\mathbf{f_1}$, and $\mathbf{g}_{W,E,k}=\mathbf{G}_{E,k}\mathbf{f_2}$, the SINR of user and eavesdroppers can be denoted as $\gamma_j=\frac{a_0|\overline{\mathbf{w}}^H\mathbf{h}_{W,j}|^2}{a_0|\overline{\mathbf{w}}^H\mathbf{g}_{W,j}|^2+1}$, $j\in\{U,(E,k)\}$.
Similar to the previous section, let $\mathbf{W}=\overline{\mathbf{w}}\overline{\mathbf{w}}^H$, $\mathbf{H}_{W,j}=\mathbf{h}_{W,j}\mathbf{h}_{W,j}^H$ and $\mathbf{G}_{W,j}=\mathbf{g}_{W,j}\mathbf{g}_{W,j}^H$. {The problem of $\text{P}_{1}$ can be transformed into 
\begin{subequations}
		\setlength{\abovedisplayskip}{3pt}
	\setlength{\belowdisplayskip}{3pt}
	\begin{alignat}{5}
	\begin{split}
	\text{P}{_{2.1}:}~ &\max_{\mathbf{ W}}~\frac{1}{ P_{tot}}\{\frac{B}{\ln2}\ln(1+\frac{a_0\text{Tr}(\mathbf{H}_{W,U}\mathbf{W})}{a_0\text{Tr}({\mathbf{G}_{W,U}}\mathbf{W})+1})\\
	&-\max_{k \in K}\frac{B}{\ln2}\ln(1+\frac{a_0\text{Tr}({\mathbf{H}_{W,E,k}}\mathbf{W})}{a_0\text{Tr}({\mathbf{G}_{W,E,k}}\mathbf{W})+1})\}
	\end{split}\nonumber\\
	s.t.~~&(9\text{b}),~(9\text{c}).\nonumber
	\end{alignat}
\end{subequations}}
{By applying Lemma 1 with SDR and introducing the variable $l_W \ge \max_{k \in K}\phi_{W,E,k}$, the problem $\text{P}_{2.1}$ can be transformed into
\begin{subequations}
		\setlength{\abovedisplayskip}{3pt}
	\setlength{\belowdisplayskip}{3pt}
	\begin{alignat}{5}
	\text{P}{_{2.2}:}~ &\max_{\mathbf{ W},t_{W,U},t_{W,E,k}}~\frac{1}{{\frac{\ln2}{B} P_{tot}}}[{\phi_{W,U}(\mathbf{W},t_{W,U})-l_{W}}]\nonumber\\
	s.t.~~&\phi_{W,E,k}(\mathbf{W},t_{W,E,k}) \le l_{W},\\
	&\phi_{W,U}(\mathbf{W},t_{W,U})-l_{W} \ge \frac{\ln2}{B}R_{th},\\
	&\mathbf{W} \succeq 0, \mathbf{W}_{mm}=1,~m=1,2,...,M,
	\end{alignat}
\end{subequations}}
where
\be
	\setlength{\abovedisplayskip}{3pt}
\setlength{\belowdisplayskip}{3pt}
\begin{split}
	\phi_{W,U}=&\ln(1+a_0\text{Tr}({\mathbf{G}_{W,U}}+\mathbf{H}_{W,U})\mathbf{W})\\
	&-t_{W,U}({a_0\text{Tr}({\mathbf{G}_{W,U}}\mathbf{W})+1})+\ln t_{W,U}+1,
\end{split}
\ee
and
\be
	\setlength{\abovedisplayskip}{3pt}
\setlength{\belowdisplayskip}{3pt}
\begin{split}
	\phi_{W,E,k}=&t_{W,E,k}(1+a_0\text{Tr}({\mathbf{G}_{W,E,k}}+\mathbf{H}_{W,E,k})\mathbf{W})\\
	&-\ln({a_0\text{Tr}({\mathbf{G}_{W,U}}\mathbf{W})+1})-\ln t_{W,E,k}-1.
\end{split}
\ee
{Since the objective function is a concave-convex fractional function, By using the Dinkelbach's method \cite{dink1}, $\text{P}_{2.2}$ can be solved by iteratively solving the following problem, given as}
\begin{subequations}
		\setlength{\abovedisplayskip}{3pt}
	\setlength{\belowdisplayskip}{3pt}
	\begin{alignat}{5}
	\text{P}{_{2.3}:}~ &\max_{\mathbf{ W},t_{W,U},t_{W,E,k}}~\phi_{W,U}(\mathbf{W},t_{W,U})-l_{W}-\frac{\ln2}{B}\eta_2^* P_{tot}\nonumber\\
	s.t.~~&\phi_{W,E,k}(\mathbf{W},t_{W,E,k}) \le l_{W},\\
	&\phi_{W,U}(\mathbf{W},t_{W,U})-l_{W} \ge \frac{\ln2}{B}R_{th},\\
	&\mathbf{W} \succeq 0, \mathbf{W}_{mm}=1,~m=1,2,...,M,
	\end{alignat}
\end{subequations}

The problem $\text{P}_{2.3}$ is a convex problem and can be solved by using the standard convex optimization tool. After the optimal $\mathbf{W}$ is obtained, $t_{W,U}$ and $t_{W,E,k}$ can be given as
$t_{W,U}^*=(a_0\text{Tr}({\mathbf{G}_{W,U}}\mathbf{W})+1)^{-1}$ and
$t_{W,E,k}^*=(1+a_0\text{Tr}({\mathbf{G}_{W,E,k}}+\mathbf{H}_{W,E,k})\mathbf{W})^{-1}$.
After obtaining $\mathbf{W}$, the $\overline{\mathbf{w}}$ can be given by eigenvalue decomposition if $\text{rank}({\mathbf{W}})=1$, otherwise, the Gaussian randomization can be used for recovering the approximate $\overline{\mathbf{w}}$ \cite{wqqirs}. The reflection coefficients can be given by
$w_m=\angle(\frac{\overline{{w}}_m}{\overline{{w}}_{M+1}}),~m=1,2,..,M$.
The overall optimization algorithm for solving $\text{P}_0$ is summarized in Algorithm 1, where $\delta$ is the threshold and $T$ is the maximum number of iterations.

{
\subsection{Convergence Analysis}
	For the convergence of the proposed algorithm, similar to \cite{irs4}, the proof is given as follows.
 Let $(\mathbf{W}^k, \mathbf{F}_1^k,\mathbf{F}_2^k)$ denote the feasible solution in the $k$th iteration, and let $J_{1.4}$ denote the objective function of $\text{P}_{1.4}$. It can be seen that for the given $\mathbf{W}^{k+1}$ and $\mathbf{W}^k$ from two iterations, one has ${J_{1.4}}({\mathbf{W}^{k + 1}},{\mathbf{F}_1^{k+ 1}},{\mathbf{F}_2^{k+ 1}})$ $\mathop \ge \limits^{(a)} {J_{1.4}}({\mathbf{W}^{k + 1}},{\mathbf{F}_1^{k}},{\mathbf{F}_2^{k}})$ $\mathop  \ge \limits^{(b)} {J_{1.4}}({\mathbf{W}^k},{\mathbf{F}_1^{k}},{\mathbf{F}_2^{k}})$, where $(a)$ holds because for the given $\mathbf{W}^{k+1}$ in Algorithm 1, $({\mathbf{F}_1^{k+ 1}},{\mathbf{F}_2^{k+ 1}})$ are the optimal solutions of problem $\text{P}_{1.4}$, and $(b)$ holds because from the objective function of $\text{P}_{2.1}$, we have $ \ln [1 + {a_0}{\rm{Tr}}({{\bf{H}}_u}{{\bf{F}}_1}{\bf{H}}_u^H{{\bf{W}}^{k + 1}}) + {\rm{Tr}}({{\bf{G}}_u}{{\bf{F}}_2}{\bf{G}}_u^H{{\bf{W}}^{k + 1}})]- {t_{W,U}}{\rm{[}}{a_0}{\rm{Tr}}({{\bf{G}}_u}{{\bf{F}}_2}{\bf{G}}_u^H{{\bf{W}}^{k + 1}}) + 1] + \ln {t_{W,U}} + 1 $
	$ = \ln [1 + {a_0}{\rm{Tr}}({\bf{H}}_u^H{{\bf{W}}^{k + 1}}{{\bf{H}}_u}{{\bf{F}}_1}) + {\rm{Tr}}({\bf{G}}_u^H{{\bf{W}}^{k + 1}}{{\bf{G}}_u}{{\bf{F}}_2})] $$- \ln ({a_0}{\rm{Tr}}({\bf{G}}_u^H{{\bf{W}}^{k + 1}}{{\bf{G}}_u}{{\bf{F}}_2}) + 1)$ $\ge $$\ln [1 + {a_0}{\rm{Tr}}({\bf{H}}_u^H{{\bf{W}}^k}{{\bf{H}}_u}{{\bf{F}}_1}) + {\rm{Tr}}({\bf{G}}_u^H{{\bf{W}}^k}{{\bf{G}}_u}{{\bf{F}}_2})] $ $- \ln ({a_0}{\rm{Tr}}({\bf{G}}_u^H{{\bf{W}}^k}{{\bf{G}}_u}{{\bf{F}}_2}) + 1)$. Similarly, one has $R_{E,k}(\mathbf{W}^{k+1},\mathbf{F}^{k}_1,\mathbf{F}^{k}_2)\le R_{E,k}(\mathbf{W}^{k},\mathbf{F}^{k}_1,\mathbf{F}^{k}_2)$. Therefore, the objective function of problem $\text{P}_{1.4}$ is non-decreasing over the iterations in the proposed algorithm, and the objective value of $\text{P}_{1.4}$ is finite due to the limited resource in the system. Thus, the proposed method is able to converge to a stationary point. A similar proof can be obtained for $\text{P}_{2.1}$.

When the obtained solutions $\mathbf{F}_1$, $\mathbf{F}_2$, and $\mathbf{W}$ are not rank-one matrices, based on Gaussian randomization, a set of $\mathbf{\zeta}_1 \sim \mathcal{CN}(0,\mathbf{F}_1)$, $\mathbf{\zeta}_2 \sim \mathcal{CN}(0,\mathbf{F}_2)$, and $\mathbf{\zeta}_3 \sim \mathcal{CN}(0,\mathbf{W})$ are generated. Then, the feasibility of $\textbf{P}_1$ is checked with $\mathbf{\zeta}_i$, $i\in \{1,2,3\}$, and the monotonicity is also checked by comparing the current results with the results from the previous iteration. Via independently generating enough feasible $\mathbf{\zeta}_i$, $i\in \{1,2,3\}$, the optimal value of problem $\textbf{P}_1$ can be approximated by the best $\mathbf{\zeta}_i$ among all random vectors with an arbitrary small bias $\epsilon >0$ \cite{rc22}.
}

The proposed method can provide a sub-optimal solution when the Gaussian randomization is applied. In Section V, we will compare the proposed method with the existing maximum ratio transmission (MRT) beamforming method to verify the superiority of our proposed scheme in terms of energy efficiency.

\vspace{-0.5cm}
\section{System Design With Imperfect CSI}
\label{imperfect}
In this section, based on the results obtained in Section \ref{perfect}, the energy efficiency maximization problem is extended into the case that the CSIs of the links from the IRS to the Eves are imperfect. The beamforming and jamming vectors and the phase shift matrix are jointly optimized to maximize the energy efficiency. 
\vspace{-0.15in}
\subsection{Problem Formulation}
By considering the imperfect CSI model between the Eves and IRS, the energy efficiency maximization problem can be formulated as 
\begin{subequations}
		\setlength{\abovedisplayskip}{3pt}
	\setlength{\belowdisplayskip}{3pt}
	\label{P3}
	\begin{alignat}{5}
	\text{P}{_3:}~ &\max_{\mathbf{f}_1,\mathbf{f}_2,\mathbf{ \Theta}}~\eta\nonumber\\
	s.t.~~&\mathbf{f}_1^{H}\mathbf{f}_1 \le P_{1,max},~\mathbf{f}_2^{H}\mathbf{f}_2 \le P_{2,max},\\
	&R_s \ge R_{th},~\Delta\mathbf{h}_{I,E,k} \in \mathcal{H}_{I,E,k}, \Delta\mathbf{g}_{I,E,k} \in\mathcal{G}_{I,E,k},\\
	&|\exp({j\theta_{m}})|=1.
	\end{alignat}
\end{subequations} 
Motivated by the method used for solving $\text{P}_1$, the problem $\text{P}_3$ can be also solved by using an alternating optimization method.

\vspace{-0.15in}
\subsection{Optimizing the Beamforming with a Given $\mathbf{\Theta}$}

In this section, we solve the problem $\text{P}_{3}$ to achieve the optimal secure transmit beamforming vector $\mathbf{f}_1$ and jamming vector $\mathbf{f}_2$ for a given $\mathbf{\Theta}$.

Let $\mathbf{H}_{B,W}=\mathbf{\Theta}\mathbf{H}_{B,I}$ and  $\mathbf{G}_{J,W}=\mathbf{\Theta}\mathbf{G}_{J,I}$ to simplify the formulas. By setting $\mathbf{H}_{B,E,k,X}=\left[ {\begin{array}{*{20}{c}}
	\mathbf{H}_{B,W}  \\
	\mathbf{h}^H_{B,E,k}  \\
	\end{array}} \right]$,  $\mathbf{G}_{J,E,k,X}=\left[ {\begin{array}{*{20}{c}}
	\mathbf{G}_{J,W}  \\
	\mathbf{g}^H_{J,E,k}  \\
	\end{array}} \right]$, and introducing ${{\mathbf{h}}_{I,E,k,X = }}\overline{\mathbf{h}}_{I,E,k,X}  + \Delta {{\mathbf{h}}_{I,E,k,X}}$, ${{\bf{g}}_{I,E,k,X}} = \overline{\bf{g}}_{I,E,k,X}  + \Delta {{\bf{g}}_{I,E,k,X}}$, where  $\overline{\mathbf{h}}_{I,E,k,X}=\left[ {\begin{array}{*{20}{c}}
	\overline{\mathbf{h}}_{I,E,k}  \\
	1  \\
	\end{array}}\right]$, $\Delta {{{\mathbf{h}}_{I,E,k,X}}}=\left[ {\begin{array}{*{20}{c}}
	\Delta {{{\mathbf{h}}_{I,E,k}}}  \\
	0  \\
	\end{array}}\right]$, $\overline{\mathbf{g}}_{I,E,k,X}=\left[ {\begin{array}{*{20}{c}}
	\overline{\mathbf{g}}_{I,E,k}  \\
	1  \\
	\end{array}}\right]$, and $\Delta {{{\mathbf{g}}_{I,E,k,X}}}=\left[ {\begin{array}{*{20}{c}}
	\Delta {{{\mathbf{g}}_{I,E,k}}}  \\
	0  \\
	\end{array}}\right]$, respectively, the SINR of Eve $k$ can be reformulated as
${\gamma _{E,k}}=\frac{{|({\mathbf{h}}_{I,E,k,X}^H{{\mathbf{H}}_{B,E,k,X}}){{\mathbf{f}}_1}{|^2}}}{{|({\mathbf{g}}_{I,E,k,X}^H{{\mathbf{G}}_{J,E,k,X}}){{\mathbf{f}}_2}{|^2} + {\sigma ^2}}}.$
{The problem $\text{P}_3$ can be transformed into

\begin{subequations}
		\setlength{\abovedisplayskip}{3pt}
	\setlength{\belowdisplayskip}{3pt}
\begin{alignat}{5}
\begin{split}
	\text{P}{_{3.1}:}~ &\max_{\mathbf{f}_1,\mathbf{f}_2}\frac{}{P_{tot}}\{\frac{B}{{\ln 2}}\ln(1 +\frac{|\overline{\mathbf{w}}^H\mathbf{H}_{U}\mathbf{f}_1|^2}{|\overline{\mathbf{w}}^H\mathbf{G}_{U}\mathbf{f}_2|^2+\sigma^2})\\
	&-{\max _{k \in K}}\frac{B}{{\ln 2}}\ln(1 +\frac{{|({\mathbf{h}}_{I,E,k,X}^H{{\mathbf{H}}_{B,E,k,X}}){{\mathbf{f}}_1}{|^2}}}{{|({\mathbf{g}}_{I,E,k,X}^H{{\mathbf{G}}_{J,E,k,X}}){{\mathbf{f}}_2}{|^2} + {\sigma ^2}}})\}
	\end{split}\nonumber\\
	s.t.~~&\mathbf{f}_1^{H}\mathbf{f}_1 \le P_{1,max},~\mathbf{f}_2^{H}\mathbf{f}_2 \le P_{2,max},\\
	\begin{split}
	&\frac{B}{{\ln 2}}\ln(1 + \frac{|\overline{\mathbf{w}}^H\mathbf{H}_{U}\mathbf{f}_1|^2}{|\overline{\mathbf{w}}^H\mathbf{G}_{U}\mathbf{f}_2|^2+\sigma^2}\\
	&-{\max _{k \in K}}\frac{B}{{\ln 2}}\ln(1 + \frac{{|({\mathbf{h}}_{I,E,k,X}^H{{\mathbf{H}}_{B,E,k,X}}){{\mathbf{f}}_1}{|^2}}}{{|({\mathbf{g}}_{I,E,k,X}^H{{\mathbf{G}}_{J,E,k,X}}){{\mathbf{f}}_2}{|^2} + {\sigma ^2}}})\\
	& \ge R_{th},~\Delta\mathbf{h}_{I,E,k,X} \in \mathcal{H}_{I,E,k}, \Delta\mathbf{g}_{I,E,k,X} \in \mathcal{G}_{I,E,k}.
	\end{split}
\end{alignat}
\end{subequations}}

Similar to the method used in Section \ref{perfect}, by defining $\mathbf{F}_1 = \mathbf{f}_1\mathbf{f}_1^H$ and $\mathbf{F}_2 = \mathbf{f}_2\mathbf{f}_2^H$, one has $\mathbf{F}_1\succeq0$, $\mathbf{F}_2\succeq0$ and $\text{rank}(\mathbf{F}_1)=\text{rank}(\mathbf{F}_2)=1$. 
The rank-1 constraint makes problem hard to be solved. By applying the SDR method to relax the rank-1 constraints \cite{sdr}, the problem $\text{P}_{3.1}$ can be transformed into
{
\begin{subequations}
		\setlength{\abovedisplayskip}{3pt}
	\setlength{\belowdisplayskip}{3pt}
	\begin{alignat}{5}
	\begin{split}
	&\text{P}{_{3.2}:}~ \max_{\mathbf{F}_1,\mathbf{F}_2} \frac{1}{P_{tot}}\big(\frac{B}{{\ln 2}}[{\ln(\text{Tr}(\overline{\mathbf{H}}_{U}\mathbf{F}_1)+\text{Tr}(\overline{\mathbf{G}}_{U}\mathbf{F}_2)+\sigma^2)}\\
	&-\ln(\text{Tr}(\overline{\mathbf{G}}_{U}\mathbf{F}_2)+\sigma^2)]\\
	&-{\max _{k \in K}}\frac{B}{{\ln 2}}[\ln( {\mathbf{h}}_{I,E,k,X}^H{\mathbf{H}}_{B,E,k,X}{{\mathbf{F}}_1}{\mathbf{H}}_{B,E,k,X}^H{\mathbf{h}}_{I,E,k,X}\\
	&+{\mathbf{g}}_{I,E,k,X}^H{{\mathbf{G}}_{J,E,k,X}}{{\mathbf{F}}_2}{{\mathbf{G}}_{J,E,k,X}^H}{\mathbf{g}}_{I,E,k,X}+ {\sigma^2})\\
	&-\ln({\mathbf{g}}_{I,E,k,X}^H{{\mathbf{G}}_{J,E,k,X}}{{\mathbf{F}}_2}{{\mathbf{G}}_{J,E,k,X}^H}{\mathbf{g}}_{I,E,k,X} + {\sigma ^2})]\big)
	\end{split}\nonumber\\
	&s.t.~~(\mathbf{F}_1, \mathbf{F}_2) \in \mathcal{F},\\
	\begin{split}
	&\frac{B}{{\ln 2}}[{\ln(\text{Tr}(\overline{\mathbf{H}}_{U}\mathbf{F}_1)+\text{Tr}(\overline{\mathbf{G}}_{U}\mathbf{F}_2)+\sigma^2)}-\ln(\text{Tr}(\overline{\mathbf{G}}_{U}\mathbf{F}_2)+\sigma^2)]\\
	&-{\max _{k \in K}}\frac{B}{{\ln 2}}[\ln( {\mathbf{h}}_{I,E,k,X}^H{\mathbf{H}}_{B,E,k,X}{{\mathbf{F}}_1}{\mathbf{H}}_{B,E,k,X}^H{\mathbf{h}}_{I,E,k,X}\\
	&+{\mathbf{g}}_{I,E,k,X}^H{{\mathbf{G}}_{J,E,k,X}}{{\mathbf{F}}_2}{{\mathbf{G}}_{J,E,k,X}^H}{\mathbf{g}}_{I,E,k,X}\\
	& + {\sigma ^2})-\ln({\mathbf{g}}_{I,E,k,X}^H{{\mathbf{G}}_{J,E,k,X}}{{\mathbf{F}}_2}{{\mathbf{G}}_{J,E,k,X}^H}{\mathbf{g}}_{I,E,k,X} + {\sigma ^2})]\\
	&\ge R_{th},\Delta\mathbf{h}_{I,E,k,X} \in \mathcal{H}_{I,E,k}, ~\Delta\mathbf{g}_{I,E,k,X} \in \mathcal{G}_{J,E,k},
	\end{split}
	\end{alignat}
\end{subequations}}
where $\mathcal{F}=\{(\mathbf{F}_1,\mathbf{F}_2)|\text{Tr}(\mathbf{F}_1)\le P_{1,max},~\text{Tr}(\mathbf{F}_2)\le P_{2,max},~\mathbf{F}_1 \succeq0,~\mathbf{F}_2 \succeq0)\}$.  Lemma 1 can be applied to solve the non-convexity caused by the second term in objective function and constraint (26b).

\begin{table}[htbp]
	\setlength{\abovedisplayskip}{3pt}
	\setlength{\belowdisplayskip}{3pt}
	\centering
	\begin{center}
		\begin{tabular}{lcl}
			\\\toprule
			{	$\textbf{Algorithm 1}$: Alternating Algorithm for Solving $\text{P}_{0}$}\\ \midrule
			\  1) \textbf{Input settings:}\\
			\ \ \ \ \ \ \ $\delta$, $R_{th},P_{1,max},P_{2,max} >0$, and $T$.\\
			\  2) \textbf{Initialization:}\\
			\ \ \ \ \ \ \ $t_{U}(0)$, $t_{E,k}(0)$, $t_{W,U}(0)$, $t_{W,E,K}(0)$, $\mathbf{w}(0)$, $\eta(0)$;\\
			\  3) \textbf{Optimization:}\\
			\ \ \ \ \ \textbf{$\pmb{\unrhd} $  for \emph{$\tau_1$}=1:T }\\
			\ \ \ \ \ \ \ \ \ \ solve $\text{P}_{1.6}$ with $(\mathbf{w}^{*}(\tau_1-1))$;\\
			\ \ \ \ \ \ \ \ \ \ obtain the solution $\mathbf{f}_1^{*}(\tau_1)$, $\mathbf{f}_2^{*}(\tau_1)$;\\
			\ \ \ \ \ \ \ \ \ \ solve $\text{P}_{2.3}$ with $(\mathbf{f}_1^{*}(\tau_1), \mathbf{f}_2^{*}(\tau_1))$;\\
			\ \ \ \ \ \ \ \ \ \ obtain the solution $\mathbf{w}^{*}(\tau_1)$;\\
			\ \ \ \ \ \ \ \ \ \ calculate energy efficiency $\eta(\tau_1)$;\\			
			\ \ \ \ \ \ \ \ \ \  \textbf{if} $\| \eta(\tau_1)-\eta(\tau_1-1)\| \le \delta$; \\
			\ \ \ \ \ \ \ \ \ \ \ \ \ \ the optimal energy efficiency $\eta^*$ is obtained;\\
			\ \ \ \ \ \ \ \ \ \  \textbf{end}\\
			\ \ \ \ \ \textbf{$\pmb{\unrhd} $ end} \\
			\  4) \textbf{Output:}\\
			\ \ \ \ \ \ \ $\{\mathbf{f}_1^{*}, \mathbf{f}_2^{*}, \mathbf{w}^*\}$ and energy efficiency $\eta^*$.\\
			\bottomrule
		\end{tabular}
	\end{center}
\end{table}

Let $x_{E,k}={\mathbf{h}}_{I,E,k,X}^H{\mathbf{H}}_{B,E,k,X}{{\mathbf{F}}_1}{\mathbf{H}}_{B,E,k,X}^H{\mathbf{h}}_{I,E,k,X}+{\mathbf{g}}_{I,E,k,X}^H{{\mathbf{G}}_{J,E,k,X}}{{\mathbf{F}}_2}{{\mathbf{G}}_{J,E,k,X}^H}{\mathbf{g}}_{I,E,k,X} + {\sigma ^2}$ and $t=t_{E,k}$, the transmit rate of Eve $k$ can be denoted as
\be
	\setlength{\abovedisplayskip}{3pt}
\setlength{\belowdisplayskip}{3pt}
\begin{aligned}
	R_{E,k}\frac{\ln2}{B}&=\ln( {\mathbf{h}}_{I,E,k,X}^H{\mathbf{H}}_{B,E,k,X}{{\mathbf{F}}_1}{\mathbf{H}}_{B,E,k,X}^H{\mathbf{h}}_{I,E,k,X}\\
	&+{\mathbf{g}}_{I,E,k,X}^H{{\mathbf{G}}_{J,E,k,X}}{{\mathbf{F}}_2}{{\mathbf{G}}_{J,E,k,X}^H}{\mathbf{g}}_{I,E,k,X} + {\sigma ^2})\\
	&-\ln({\mathbf{g}}_{I,E,k,X}^H{{\mathbf{G}}_{J,E,k,X}}{{\mathbf{F}}_2}{{\mathbf{G}^H}_{J,E,k,X}}{\mathbf{g}}_{I,E,k,X} + {\sigma ^2})\\
	&=\min_{t_{E,k\ge 0}}\phi_{E,k}(\mathbf{F}_1,\mathbf{F}_2,t_{E,k}),
\end{aligned}
\ee
where $\phi ({t_{E,k}}) ={t_{E,k}}({\mathbf{h}}_{I,E,k,X}^H{\mathbf{H}}_{B,E,k,X}{{\mathbf{F}}_1}{\mathbf{H}}_{B,E,k,X}^H{\mathbf{h}}_{I,E,k,X}$ $+{\mathbf{g}}_{I,E,k,X}^H{{\mathbf{G}}_{J,E,k,X}}{{\mathbf{F}}_2}{{\mathbf{G}}_{J,E,k,X}^H}{\mathbf{g}}_{I,E,k,X} + {\sigma ^2})$ $+\ln({\bf{g}}_{I,E,k,X}^H{{\bf{G}}_{J,E,k,X}}{{\bf{F}}_2}{{\bf{G}}^H_{J,E,k,X}}{\bf{g}}_{I,E,k,X} + {\sigma ^2})- \ln ({t_{E,k}}) - 1$.
Therefore, the problem $\text{P}_{3.2}$ can be transformed into
\begin{subequations}
		\setlength{\abovedisplayskip}{3pt}
	\setlength{\belowdisplayskip}{3pt}
	\begin{alignat}{5}
	\begin{split}
	\text{P}{_{3.3}:}~ &\max_{\mathbf{F}_1,\mathbf{F}_2}\max_{t_U}\frac{1}{\frac{\ln2}{B}{P_{tot}}}[\phi_U(\mathbf{F}_1,\mathbf{F}_2,t_U)\\
	&-\min_{t_{E,k}}\phi_{E,k}(\mathbf{F}_1,\mathbf{F}_2,t_{E,k})]
	\end{split}\nonumber\\
	s.t.~~&(\mathbf{F}_1, \mathbf{F}_2) \in \mathcal{F},\\
	\begin{split}
	&\max_{t_U}\phi_U(\mathbf{F}_1,\mathbf{F}_2,t_U)-\min_{t_{E,k}}\phi_{E,k}(\mathbf{F}_1,\mathbf{F}_2,t_{E,k}) \ge R_{th},\\
		&\Delta\mathbf{h}_{I,E,k,X} \in \mathcal{H}_{I,E,k}, \Delta\mathbf{g}_{I,E,k,X} \in \mathcal{G}_{J,E,k}.
	\end{split}
	\end{alignat}
\end{subequations}
By using Sion's minimax theorem \cite{sion}, and introducing the slack variable $l \ge \max_{k \in K} \phi_{E,k}$, the problem $\text{P}_{3.3}$ can be further transformed into
\begin{subequations}
		\setlength{\abovedisplayskip}{3pt}
	\setlength{\belowdisplayskip}{3pt}
	\begin{alignat}{5}
\begin{split}
\text{P}_{3.4}&~~\max_{\mathbf{F}_1, \mathbf{F}_2, t_U, t_{E,k}}\frac{\phi_U({\mathbf{F}_1,\mathbf{F}_2, }t_U)-l}{\frac{\ln2}{B} (\text{Tr}(\mathbf{F}_1+\mathbf{F}_2)+P_{BS}+P_{G}+P_{IRS})}
\end{split}\nonumber\\
\text{s.t.}&~~(\mathbf{F}_1, \mathbf{F}_2) \in \mathcal{F},t_U, t_{E,k} \ge 0,\\
&\phi_U({\mathbf{F}_1,\mathbf{F}_2, }t_U)-l \ge R_{th}\frac{\ln2}{B},\\
&\phi_{E,k}({\mathbf{F}_1,\mathbf{F}_2, }t_{E,k})\le l,\\
&~\Delta\mathbf{h}_{I,E,k,X} \in \mathcal{H}_{I,E,k},~\Delta\mathbf{g}_{I,E,k,X}\in \mathcal{G}_{I,E,k}.
\end{alignat}
\end{subequations}
However, the problem is still difficult to be solved due to the uncertainty of the CSI from the IRS to the Eves. We introduce the slack variable $\psi_{B,E,k}$, and $\psi_{J,E,k}$ to deal with this uncertainty.
\begin{subequations}
		\setlength{\abovedisplayskip}{3pt}
	\setlength{\belowdisplayskip}{3pt}
\begin{alignat}{5}
&{\bf{h}}_{I,E,k,X}^H{{\bf{H}}_{B,E,k,X}}{{\bf{F}}_1}{{\bf{H}}_{B,E,k,X}^H}{\bf{h}}_{I,E,k,X} \le {\psi _{B,E,k}},\\
&{\bf{g}}_{I,E,k,X}^H{{\bf{G}}_{J,E,k,X}}{{\bf{F}}_2}{{\bf{G}}_{J,E,k,X}^H}{\bf{g}}_{I,E,k,X} \ge {\psi _{J,E,k}}.
\end{alignat}
\end{subequations}
Then we have $\phi_{E,k}\le t_{E,k}(\psi_{B,E,k}+\psi_{J,E,k}+\sigma^2)-\ln(\psi_{J,E,k}+\sigma^2)-\ln(t_{E,k})-1$. {The problem $\text{P}_{3.4}$ can be transformed into
\begin{subequations}
		\setlength{\abovedisplayskip}{3pt}
	\setlength{\belowdisplayskip}{3pt}
	\begin{alignat}{5}
	\begin{split}
	\text{P}_{3.5}&~~\max_{\mathbf{F}_1, \mathbf{F}_2, t_U, t_{E,k},\psi_{B,E,k},\psi_{J,E,k}}
	\frac{1}{\frac{\ln2}{B} P_{tot}}[{\phi_U({\mathbf{F}_1,\mathbf{F}_2, }t_U)-l}]
	\end{split}\nonumber\\
	\text{s.t.}&~~(29\text{a}),(29\text{b}),(30\text{a}),(30\text{b}),\nonumber\\
	&t_{E,k}(\psi_{B,E,k}+\psi_{J,E,k}+\sigma^2)-\ln(\psi_{J,E,k}+\sigma^2)\\
	&-\ln(t_{E,k})-1 \le l, \forall k.
	\end{alignat}
\end{subequations}}

$\text{P}_{3.5}$ can be solved by alternately solving $(t_U,t_{E,k})$ and $(\mathbf{F}_1,\mathbf{F}_1)$. First, with the given $(t_U^*,t_{E,k}^*)$, to solve the problem $\text{P}_{3.5}$ for $(\mathbf{F}_1,\mathbf{F}_1)$, the $\mathcal{S}$-Procedure is applied.

$\mathbf{Lemma~2}$: Let $f_i(\mathbf{z})=\mathbf{z}^H\mathbf{A}_i\mathbf{z}+2\Re({\mathbf{b}_i^H\mathbf{z}})+c_i, i\in\{1,2\}$, where $\mathbf{z} \in \mathbb{C}^{M\times1}$, $\mathbf{A}_i \in \mathbb{C}^{M\times M}$, $\mathbf{b_i} \in \mathbb{C}^{M\times1}$, and $c_i \in \mathbb{R}$. Then, the expression $f_1(\mathbf{z})\le 0 \Rightarrow f_2(\mathbf{z})\le 0$ holds if and only if there exists a $\lambda\ge 0$ such that
\be
	\setlength{\abovedisplayskip}{3pt}
\setlength{\belowdisplayskip}{3pt}
\lambda \left[ {\begin{array}{*{20}{c}}
	{{{\bf{A}}_1}} & {{{\bf{b}}_1}}  \\
	{{\bf{b}}_1^H} & {{c_1}}  \\
	\end{array}} \right] - \left[ {\begin{array}{*{20}{c}}
	{{{\bf{A}}_2}} & {{{\bf{b}}_2}}  \\
	{{\bf{b}}_2^H} & {{c_2}}  \\
	\end{array}} \right] \succeq 0,
\ee
which assumes that there exists a vector $\overline{\mathbf{z}}$ such that $f(\overline{\mathbf{z}})< 0$ \cite{rfzar1}.
By applying Lemma 2, let $\overline{\bf{h}}_{E,k,X}={{\bf{H}}^H_{B,E,k,X}}\overline {{{\bf{h}}}}{{_{I,E,k,X}}}$ and $\overline{\bf{g}}_{E,k,X}={{\bf{G}}^H_{J,E,k,X}}\overline {{{\bf{g}}}}{{_{I,E,k,X}}}$, the constraint (30a)-(30b) can be transformed into (33) and (34).
	\begin{figure*}
\be
	\setlength{\abovedisplayskip}{3pt}
\setlength{\belowdisplayskip}{3pt}
\left[ {\begin{array}{*{20}{c}}
	{{\lambda _{B,E,k}}\mathbf{I} - {{\bf{H}}_{B,E,k,X}}{{\bf{F}}_1}{{\bf{H}}^H_{B,E,k,X}}} & { - {{\bf{H}}_{B,E,k,X}}{{\bf{F}}_1}{{\bf{H}}^H_{B,E,k,X}}\overline {{{\bf{h}}}}{{_{I,E,k,X}}} }  \\
	{ - {{\overline {{{\bf{h}}}}{{^H_{I,E,k,X}}} }}{{\bf{H}}_{B,E,k,X}}{{\bf{F}}_1}{{\bf{H}}^H_{B,E,k,X}}} & {  {\psi _{B,E,k}}- {\lambda _{B,E,k}}{\xi ^2_{I,E,k}} - \overline{\bf{h}}_{E,k,X}^H{{\bf{F}}_1}\overline{\bf{h}}_{E,k,X}  }  \\
	\end{array}} \right]\succeq0,
\ee
\be
\left[ {\begin{array}{*{20}{c}}
	{{\lambda _{J,E,k}}\mathbf{I} + {{\bf{G}}_{J,E,k,X}}{{\bf{F}}_2}{{\bf{G}}^H_{J,E,k,X}}} & {{{\bf{G}}_{J,E,k,X}}{{\bf{F}}_2}{{\bf{G}}^H_{J,E,k,X}}\overline {{{\bf{g}}}}{{_{I,E,k,X}}} }  \\
	{{{\overline {{{\bf{g}}}}{{^H_{I,E,k,X}}} }}{{\bf{G}}_{J,E,k,X}}{{\bf{F}}_2}{{\bf{G}}^H_{J,E,k,X}}} & {{-\lambda _{J,E,k}}\xi _{J,E,k}^2- {\psi _{J,E,k}} + \overline{\bf{g}}_{E,k,X}^H{{\bf{F}}_2}\overline{\bf{g}}_{E,k,X}  }  \\
	\end{array}} \right]\succeq0.
\ee
\end{figure*}

{Then, similar to the  previous section, by introducing the variable $\eta_3^*$, the optimization problem $\text{P}_{3.5}$ for $\mathbf{F}_1$ and $\mathbf{F}_2$ based on $t_U$ and $t_{E,k}$ can be given as}

\begin{subequations}
		\setlength{\abovedisplayskip}{3pt}
	\setlength{\belowdisplayskip}{3pt}
	\begin{alignat}{5}
	\begin{split}
	&\text{P}_{3.6}~~\max_{\mathbf{F}_1, \mathbf{F}_2, \psi_{B,E,k},\psi_{J,E,k},\lambda_{B,E,k},\lambda_{J,E,k}} \phi_U({\mathbf{F}_1,\mathbf{F}_2, }t_U)\\
	&-l-\frac{\ln2}{B}\eta^* P_{tot}
	\end{split}\nonumber\\
	&\text{s.t.}~~(29\text{a}),(29\text{b}),(30\text{a}),(33), (34).\nonumber
	\end{alignat}
\end{subequations}
The problem $\text{P}_{3.6}$ is a convex problem since the  objective function and the constraints are all convex. It can be solved by using a standard convex optimization tool. 

After $\mathbf{F_1}$ and $\mathbf{F_2}$ are obtained, if $\text{rank}(\mathbf{F_1})=\text{rank}(\mathbf{F_2})=1$, $\mathbf{f_1}$ and $\mathbf{f_2}$ can be obtained from $\mathbf{F_1}=\mathbf{f_1}\mathbf{f_1^H}$ and $\mathbf{F_2}=\mathbf{f_2}\mathbf{f_2^H}$ by applying the eigenvalue decomposition. Otherwise, the Gaussian randomization can be used for recovering the approximate $\mathbf{f_1}$ and $\mathbf{f_2}$. 
After $\mathbf{f_1}$ and $\mathbf{f_2}$ are obtained, according to Lemma 1, the optimal value of $t_U$ can be achieved when
\be
	\setlength{\abovedisplayskip}{3pt}
\setlength{\belowdisplayskip}{3pt}
t_U^*=(\text{Tr}(\overline{\mathbf{G}}_{U}\mathbf{F}_2)+\sigma^2)^{-1}.
\ee
To optimize $t_{E,k}$, the following problem should be solved.
\be
	\setlength{\abovedisplayskip}{3pt}
\setlength{\belowdisplayskip}{3pt}
\begin{split}
\max_{t_{E,k}}&-{t_{E,k}}(\max_{\Delta\mathbf{ h}_{I,E,k,X}}{\mathbf{h}}_{I,E,k,X}^H{\mathbf{H}}_{B,E,k,X}{{\mathbf{F}}_1}{\mathbf{H}}_{B,E,k,X}^H{\mathbf{h}}_{I,E,k,X}\\
&+\min_{\Delta\mathbf{g}_{J,E,k,X}}{\mathbf{g}}_{I,E,k,X}^H{{\mathbf{G}}_{J,E,k,X}}{{\mathbf{F}}_2}{{\mathbf{G}}_{J,E,k,X}^H}{\mathbf{g}}_{I,E,k,X}\\
& + {\sigma ^2})+ \ln ({t_{E,k}}) + 1.
\end{split}
\label{l2}
\ee
This needs to first solve the following problems, given as, 
\begin{subequations}
		\setlength{\abovedisplayskip}{3pt}
	\setlength{\belowdisplayskip}{3pt}
	\begin{alignat}{5}
	\Gamma_{1,k}=\max_{\Delta\mathbf{h}_{I,E,k,X}}&{\mathbf{h}}_{I,E,k,X}^H{\mathbf{H}}_{B,E,k,X}{{\mathbf{F}}_1}{\mathbf{H}}_{B,E,k,X}^H{\mathbf{h}}_{I,E,k,X}\\
	&\text{s.t.}~ \Delta\mathbf{h}_{I,E,k,X}^H\Delta\mathbf{h}_{I,E,k,X} \le \xi_{I,E,k}^2,
\end{alignat}	\label{g1} 
\end{subequations}
and
\begin{subequations}
		\setlength{\abovedisplayskip}{3pt}
	\setlength{\belowdisplayskip}{3pt}
	\begin{alignat}{5}
	\Gamma_{2,k}=\min_{\Delta\mathbf{g}_{J,E,k,X}}&{\mathbf{g}}_{I,E,k,X}^H{{\mathbf{G}}_{J,E,k,X}}{{\mathbf{F}}_2}{{\mathbf{G}}_{J,E,k,X}^H}{\mathbf{g}}_{I,E,k,X} + {\sigma ^2}\\
	&\text{s.t.}~ \Delta\mathbf{g}_{I,E,k,X}^H\Delta\mathbf{g}_{I,E,k,X} \le \xi_{J,E,k}^2.
	\end{alignat}	\label{g2} 
\end{subequations}
For notational simplification, we denote $ \mathbf{H}_{I,E,k,X} \mathbf{F}_1 \mathbf{H}_{I,E,k,X}^H=\mathbf{F}_{1,k,X}$. Then, the Lagrangian function of problem (\ref{g1}) can be given as
\be
	\setlength{\abovedisplayskip}{3pt}
\setlength{\belowdisplayskip}{3pt}
\begin{split}
L_{1,k}&=(\overline{\mathbf{h}}_{I,E,k,X}^H+\Delta\mathbf{h}_{I,E,k,X}^H)\mathbf{F}_{1,k,X}(\overline{\mathbf{h}}_{I,E,k,X}+\Delta\mathbf{h}_{I,E,k,X})\\
&+\mu_{1,k}(\xi_{I,E,k}^2-\Delta\mathbf{h}_{I,E,k,X}^H\Delta\mathbf{h}_{I,E,k,X}),
\end{split}
\ee
where $\mu_{1,k}$ is the Lagrange multiplier. $L_{1,k}$ is convex respect to $\Delta\mathbf{h}_{I,E,k,X}$. The  Karush-Kuhn-Tucker (KKT) condition can be applied to solve this problem. Thus, one has
\be
	\setlength{\abovedisplayskip}{3pt}
\setlength{\belowdisplayskip}{3pt}
\begin{split}
\Gamma_{1,k}&=\text{tr}[\mathbf{F}_{1,k,X}(\overline{\mathbf{h}}_{I,E,k,X}\overline{\mathbf{h}}_{I,E,k,X}^H+\xi_{I,E,k}^2\mathbf{I}\\
&+2\xi_{I,E,k}\sqrt{\frac{\overline{\mathbf{h}}_{I,E,k,X}^H\mathbf{F}_{1,k,X}\mathbf{h}_{I,E,k,X}}{\text{tr}(\mathbf{F}_{1,k,X})}}\mathbf{I})].
\end{split}
\ee
	\setlength{\abovedisplayskip}{3pt}
\setlength{\belowdisplayskip}{3pt}
Similarly, letting ${{\mathbf{G}}_{J,E,k,X}}{{\mathbf{F}}_2}{{\mathbf{G}}_{J,E,k,X}^H}=\mathbf{F}_{2,k,X}$, one has
\be
\begin{split}
\Gamma_{2,k}&=\text{tr}[\mathbf{F}_{2,k,X}(\overline{\mathbf{g}}_{I,E,k,X}\overline{\mathbf{g}}_{I,E,k,X}^H+\xi_{J,E,k}^2\mathbf{I}\\
&-2\xi_{J,E,k}\sqrt{\frac{\overline{\mathbf{g}}_{I,E,k,X}^H\mathbf{F}_{2,k,X}\mathbf{g}_{I,E,k,X}}{\text{tr}(\mathbf{F}_{2,k,X})}}\mathbf{I})].
\end{split}
\ee
The closed-form expression for solution for $t_{E,k}$ can be given as
\be
	\setlength{\abovedisplayskip}{3pt}
\setlength{\belowdisplayskip}{3pt}
t_{E,k}^*=(\Gamma_{1,k}+\Gamma_{2,k}+\sigma^2)^{-1}.
\ee

Thus, the problem $\text{P}_{3.1}$ can be solved by alternately updating $(t_U,t_{E,k})$ and $(\mathbf{f_1},\mathbf{f_2})$, which is summarized at Algorithm 2. 
\vspace{-0.15in}
\subsection{Optimizing $\mathbf{w}$ with Given $(\mathbf{f}_1,\mathbf{f}_2)$}
After obtaining $\mathbf{f}_1$ and $\mathbf{f}_2$, by setting
${{\bf{H}}_{B,E,k,F}} = \left[ {\begin{array}{*{20}{c}}
{\text{diag}({{\bf{H}}_{B,I}}{{\bf{f}}_1})^H} & {}  \\
{} & {{\bf{h}}_{B,E,k}^H{{\bf{f}}_1}}  \\
\end{array}} \right]$,
and ${{\bf{G}}_{J,E,k,F}} = \left[ {\begin{array}{*{20}{c}}
{\text{diag}({{\bf{G}}_{J,I}}{{\bf{f}}_2})^H} & {}  \\
{} & {{\bf{g}}_{J,E,k}^H{{\bf{f}}_2}}  \\
\end{array}} \right]$, the SINR of Eve $k$ can be given as
\be
	\setlength{\abovedisplayskip}{3pt}
\setlength{\belowdisplayskip}{3pt}
\gamma_{E,k}=\frac{{{\bf{h}}_{I,E,k,X}^H{{\bf{H}}_{B,E,k,F}}\mathbf{{W}} {{\bf{H}}^H_{B,E,k,F}}{\bf{h}}_{I,E,k,X}}}{{{\bf{g}}_{I,E,k,X}^H{{\bf{G}}_{J,E,k,F}}\mathbf{{W}} {{\bf{G}}^H_{J,E,k,F}}{\bf{g}}_{I,E,k,X} + {\sigma ^2}}},~~\forall k \in K,
\ee
where $\mathbf{{W}}=\overline{\mathbf{w}}\overline{\mathbf{w}}^H$, $\mathbf{{W}}\succeq 0$, and $\text{Rank}(\mathbf{{W}})=1$.
{The problem of $\text{P}_{3}$ can be transformed into 
\begin{subequations}
		\setlength{\abovedisplayskip}{3pt}
	\setlength{\belowdisplayskip}{3pt}
	\begin{alignat}{5}
\begin{split}
	\text{P}{_{4.1}:}~ &\max_{\mathbf{{W}}}~\frac{1}{P_{tot}}\big[\frac{B}{\ln2}\ln(1+\frac{\text{Tr}(\mathbf{H}_{W,U}\mathbf{{W}})}{\text{Tr}({\mathbf{G}_{W,U}}\mathbf{{W}})+\sigma^2})\\
	&-\max_{k \in K}\frac{B}{\ln2}\ln(1\\
	&+\frac{{{\bf{h}}_{I,E,k,X}^H{{\bf{H}}_{B,E,k,F}}\mathbf{{W}} {{\bf{H}}^H_{B,E,k,F}}{\bf{h}}_{I,E,k,X}}}{{{\bf{g}}_{I,E,k,X}^H{{\bf{G}}_{J,E,k,F}}\mathbf{{W}} {{\bf{G}}^H_{J,E,k,F}}{\bf{g}}_{I,E,k,X} + {\sigma ^2}}})\big]
	\end{split}\nonumber\\
	s.t.~~&\mathbf{{W}}\succeq 0,\text{Rank}(\mathbf{{W}})=1,\mathbf{{W}}_{m,m}=1,~m=1,2,...,M,\\
	&R_s \ge R_{th},~\Delta\mathbf{h}_{I,E,k} \in \mathcal{H}_{I,E,k}, \Delta\mathbf{g}_{I,E,k} \in \mathcal{G}_{I,E,k}.
\end{alignat}
\end{subequations}}
Similar to the previous section, by applying Lemma 1 with SDR and introducing the variable $t_{W,U}$, $t_{W,E,K}$, and $l_W \ge \max_{k \in K}\phi_{W,E,k}$, {the problem $\text{P}_{4.1}$ can be transformed into
\begin{subequations}
		\setlength{\abovedisplayskip}{3pt}
	\setlength{\belowdisplayskip}{3pt}
	\begin{alignat}{5}
	\text{P}{_{4.2}:}~ &\max_{\mathbf{{W}},t_{W,U},t_{W,E,k}}~\frac{1}{\frac{\ln2}{B} P_{tot}}[{\phi_{W,U}(\mathbf{{W}},t_{W,U})-l_{W}}]\nonumber\\
	s.t.~~	&\phi_{W,U}(\mathbf{{W}},t_{W,U})-l_{W} \ge \frac{\ln2}{B}R_{th},\\
	&\phi_{W,E,k}(\mathbf{{W}},t_{W,E,k}) \le l_{W},\\
	&\Delta\mathbf{h}_{I,E,k} \in \mathcal{H}_{I,E,k}, \Delta\mathbf{g}_{I,E,k} \in \mathcal{G}_{I,E,k},\\
	&\mathbf{{W}} \succeq 0, \mathbf{{W}}_{mm}=1,~m=1,2,...,M,\\
	&t_{W,U} >0, t_{W,E,k} >0,~k=1,...,K,
	\end{alignat}
\end{subequations}}
where
\be
	\setlength{\abovedisplayskip}{3pt}
\setlength{\belowdisplayskip}{3pt}
\begin{split}
	\phi_{W,E,k}=&t_{W,E,k}({{{\bf{g}}^H_{I,E,k,X}{{\bf{G}}_{J,E,k,F}}\mathbf{{W}} {{\bf{G}}^H_{J,E,k,F}}{\bf{g}}_{I,E,k,X}^{} + {\sigma ^2}}}\\
	&+{{{\bf{h}}^H_{I,E,k,X}{{\bf{H}}_{B,E,k,F}}\mathbf{{W}} {{\bf{H}}^H_{B,E,k,F}}{\bf{h}}_{I,E,k,X}^{}}})\\
	&-\ln({{{\bf{g}}^H_{I,E,k,X}{{\bf{G}}_{J,E,k,F}}\mathbf{{W}} {{\bf{G}}^H_{J,E,k,F}}{\bf{g}}_{I,E,k,X} + {\sigma ^2}}})\\
	&-\ln t_{W,E,k}-1.
\end{split}
\ee
To solve the uncertainty channel constraints, we introduce the variables $\psi_{B,E,k}^{W}$ and $\psi_{J,E,k}^{W}$, which are given as
\be
	\setlength{\abovedisplayskip}{3pt}
\setlength{\belowdisplayskip}{3pt}
\begin{aligned}
&{{{\bf{h}}_{I,E,k,X}^H{{\bf{H}}_{B,E,k,F}}\mathbf{{W}} {{\bf{H}}^H_{B,E,k,F}}{\bf{h}}_{I,E,k,X}}} \le \psi_{B,E,k}^W,\\
&{\bf{g}}_{I,E,k,X}^H{{\bf{G}}_{J,E,k,F}}\mathbf{{W}} {{\bf{G}}^H_{J,E,k,F}}{\bf{g}}_{I,E,k,X} \ge \psi_{J,E,k}^W.
\end{aligned}
\ee
{Thus, problem $\text{P}_{4.2}$ can be transformed into
\begin{subequations}
		\setlength{\abovedisplayskip}{3pt}
	\setlength{\belowdisplayskip}{3pt}
	\begin{alignat}{5}
	\text{P}{_{4.3}:}~ &\max_{\mathbf{{W}},t_{W,U},t_{W,E,k},\psi_{B,E,k}^W,\psi_{J,E,k}^W}~\frac{[{\phi_{W,U}(\mathbf{{W}},t_{W,U})-l_{W}}]}{\frac{\ln2}{B} P_{tot}}\nonumber\\
	s.t.~~&(46\text{a}),(46\text{c}),(46\text{d}),\nonumber\\
	&t_{W,E,k}( \psi_{J,E,k}^W+ {\sigma ^2}+\psi_{B,E,k}^W)-\ln(\psi_{J,E,k}^W + {\sigma ^2})\\
	&-\ln(t_{W,E,k})-1 \le l_{W},\\
	&{{{\bf{h}}_{I,E,k,X}^H{{\bf{H}}_{B,E,k,F}}\mathbf{{W}} {{\bf{H}}^H_{B,E,k,F}}{\bf{h}}_{I,E,k,X}}} \le \psi_{B,E,k}^W,\\ \nonumber
	&\Delta\mathbf{h}_{I,E,k} \in \mathcal{H}_{I,E,k},\\
	&{\bf{g}}_{I,E,k,X}^H{{\bf{G}}_{J,E,k,F}}\mathbf{{W}} {{\bf{G}}^H_{J,E,k,F}}{\bf{g}}_{I,E,k,X} \ge \psi_{J,E,k}^W,\\ \nonumber
	&\Delta\mathbf{g}_{I,E,k} \in \mathcal{G}_{I,E,k}.
	\end{alignat}
\end{subequations}}
By using Lemma 2, letting $\overline{\bf{h}}_{E,k,F}={\bf{H}}_{B,E,k,F}^H{{\overline {\bf{h}} }_{I,E,k,X}}$ $\overline{\bf{g}}_{E,k,F}={\bf{G}}_{J,E,k,F}^H{{\overline {\bf{g}} }_{J,E,k,X}}$, and introducing the variable $\eta_4^*$, the problem $\text{P}_{4.3}$ can be transformed as $\text{P}{_{4.4}}$.
\begin{figure*}
			\setlength{\abovedisplayskip}{3pt}
	\setlength{\belowdisplayskip}{3pt}
\begin{subequations}
		\setlength{\abovedisplayskip}{3pt}
	\setlength{\belowdisplayskip}{3pt}
	\begin{alignat}{5}
	&\text{P}{_{4.4}:}~ \max_{\mathbf{{W}},t_{W,U},t_{W,E,k},\psi_{B,E,k}^W,\psi_{J,E,k}^W,\lambda_{B,E,k}^W,\lambda_{J,E,k}^W}~\phi_{W,U}(\mathbf{{W}},t_{W,U})-l_{W}-\frac{\ln2}{B}\eta_4^* P_{tot}\nonumber\\
	&s.t.~~(46\text{a}),(46\text{c}),(46\text{d}),(49\text{a}),\nonumber\\
	&\left[ {\begin{array}{*{20}{c}}
		{{\lambda _{B,E,k}^W}\mathbf{I} - {{\bf{H}}_{B,E,k,F}}\mathbf{{W}} {\bf{H}}_{B,E,k,F}^H} & {{-{\bf{H}}_{B,E,k,F}}\mathbf{{W}} {\bf{H}}_{B,E,k,F}^H{{\overline {\bf{h}} }_{I,E,k,X}}}  \\
		{-\overline {\bf{h}} _{I,E,k,X}^H{{\bf{H}}_{B,E,k,F}}\mathbf{{W}} {\bf{H}}_{B,E,k,F}^H} & { - {\lambda _{B,E,k}^W}\xi _{I,E,k}^2 + {\psi _{B,E,k}^W} - \overline{\bf{h}}_{E,k,F}^H{\bf{W}} \overline{\bf{h}}_{E,k,F} }
		\end{array}} \right]\succeq 0,\\
	&\left[ {\begin{array}{*{20}{c}}
		{{\lambda _{J,E,k}^W}\mathbf{I} + {{\bf{G}}_{J,E,k,F}}\mathbf{{W}} {\bf{G}}_{J,E,k,F}^H} & {  {{\bf{G}}_{J,E,k,F}}\mathbf{{W}} {\bf{G}}_{J,E,k,F}^H{{\overline {\bf{g}} }_{I,E,k,X}}}  \\
		{  \overline {\bf{g}} _{I,E,k,X}^H{{\bf{G}}_{J,E,k,F}}\mathbf{{W}} {\bf{G}}_{J,E,k,F}^H} & { - {\lambda _{J,E,k}^W}\xi _{J,E,k}^2 - {\psi _{J,E,k}^W} + \overline{\bf{g}}_{E,k,F}^H\mathbf{{W}} \overline{\bf{g}}_{E,k,F} }  \\
		\end{array}} \right] \succeq0.
	\end{alignat}
\end{subequations}
\end{figure*}
The problem $\text{P}_{4.4}$ is a convex problem with respect to ${\bf{W}}$ or $(t_{W,U}, t_{W,E,k})$ when other variables are fixed and can be solved by using a standard   convex optimization tool. 
After obtaining $\mathbf{{W}}$,  $\overline{\mathbf{w}}$ can be given by eigenvalue decomposition if $\text{rank}(\mathbf{{W}})=1$; otherwise, the Gaussian randomization can be used for recovering the approximate $\mathbf{{w}}$. 
With the optimal $\mathbf{{W}}$, one has 
\be
	\setlength{\abovedisplayskip}{3pt}
\setlength{\belowdisplayskip}{3pt}
t_{W,U}^*=(\text{Tr}({\mathbf{G}_{W,U}}\mathbf{{W}})+\sigma^2)^{-1}.
\ee
And $t_{W,E,k}$ can obtained by solving the following problems.
\be
	\setlength{\abovedisplayskip}{3pt}
\setlength{\belowdisplayskip}{3pt}
\begin{split}
&\max_{t_{W,E,k}}-t_{W,E,k}(\min_{\Delta\mathbf{g}_{J,E,k,X}}{\bf{g}}^H_{I,E,k,X}{{\bf{G}}_{J,E,k,F}}\mathbf{{W}} {{\bf{G}}^H_{J,E,k,F}}{\bf{g}}_{I,E,k,X}\\ &+\max_{\Delta\mathbf{h}_{I,E,k,X}}{{{\bf{h}}^H_{I,E,k,X}{{\bf{H}}_{B,E,k,F}}\mathbf{{W}} {{\bf{H}}^H_{B,E,k,F}}{\bf{h}}_{I,E,k,X}}}+ {\sigma ^2})\\
&+\ln t_{W,E,k}+1.
\end{split}
\ee
Let ${{\bf{H}}_{B,E,k,F}}\mathbf{{W}} {{\bf{H}}^H_{B,E,k,F}}={\bf{W}}_{B,k,X}$, and ${{\bf{G}}_{J,E,k,F}}\mathbf{{W}} {{\bf{G}}^H_{J,E,k,F}}={\bf{W}}_{J,k,X}$.
The solution for $t_{W,E,k}$ can be given as
\be
	\setlength{\abovedisplayskip}{3pt}
\setlength{\belowdisplayskip}{3pt}
t_{W,E,k}^*=(\Gamma_{W,1,k}+\Gamma_{W,2,k}+\sigma^2)^{-1},
\ee
$\Gamma_{W,1,k}$ and $\Gamma_{W,2,k}$ are respectively given as
\be
	\setlength{\abovedisplayskip}{3pt}
\setlength{\belowdisplayskip}{3pt}
\begin{split}
\Gamma_{W,1,k}&=\text{tr}[\mathbf{W}_{B,k,X}(\overline{\mathbf{h}}_{I,E,k,X}\overline{\mathbf{h}}_{I,E,k,X}^H+\xi_{I,E,k}^2\mathbf{I}\\
&+2\xi_{I,E,k}\sqrt{\frac{\overline{\mathbf{h}}_{I,E,k,X}^H\mathbf{W}_{B,k,X}\mathbf{h}_{I,E,k,X}}{\text{tr}(\mathbf{W}_{B,k,X})}}\mathbf{I})],
\end{split}
\ee
and 
\be
	\setlength{\abovedisplayskip}{3pt}
\setlength{\belowdisplayskip}{3pt}
\begin{split}
\Gamma_{W,2,k}&=\text{tr}[\mathbf{W}_{J,k,X}(\overline{\mathbf{g}}_{I,E,k,X}\overline{\mathbf{g}}_{I,E,k,X}^H+\xi_{J,E,k}^2\mathbf{I}\\
&-2\xi_{J,E,k}\sqrt{\frac{\overline{\mathbf{g}}_{I,E,k,X}^H\mathbf{W}_{J,k,X}\mathbf{g}_{I,E,k,X}}{\text{tr}(\mathbf{W}_{J,k,X})}}\mathbf{I})].
\end{split}
\ee
The overall optimization algorithm for solving $\text{P}_3$ is summarized in Algorithm 2, where $\delta$ is the threshold and $T$ is the maximum number of iterations.

\begin{table}
		\centering
	\begin{center}
		\begin{tabular}{lcl}
			\\\toprule
		{	$\textbf{Algorithm 2}$: Alternating Algorithm for Solving $\text{P}_{3}$}\\ \midrule
			\  1) \textbf{Input settings:}\\
			\ \ \ \ \ \ \  $\delta$, $R_{th}, P_{1,max},P_{2,max} >0$, and $T$.\\
			\  2) \textbf{Initialization:}\\
			\ \ \ \ \ \ \ $t_{U}(0)$, $t_{E,k}(0)$, $t_{W,U}(0)$, $t_{W,E,K}(0)$, $\mathbf{w}(0),\eta(0)$;\\
			\  3) \textbf{Optimization:}\\
			\ \ \ \ \ \textbf{$\pmb{\unrhd} $  for \emph{$\tau_2$}=1:T }\\
			\ \ \ \ \ \ \ \ \ \ solve $\text{P}_{3.1}$ with $(\mathbf{w}^{*}(\tau_2-1))$;\\
			\ \ \ \ \ \ \ \ \ \ obtain the solution $\mathbf{f}_1^{*}(\tau_2)$, $\mathbf{f}_2^{*}(\tau_2)$;\\
			\ \ \ \ \ \ \ \ \ \ solve $\text{P}_{4.1}$ with $(\mathbf{f}_1^{*}(\tau_2), \mathbf{f}_2^{*}(\tau_2))$;\\
			\ \ \ \ \ \ \ \ \ \ obtain the solution $\mathbf{w}^{*}(\tau_2)$\\
			\ \ \ \ \ \ \ \ \ \ calculate energy efficiency $\eta(\tau_2)$;\\			
			\ \ \ \ \ \ \ \ \ \  \textbf{if} $\| \eta(\tau_2)-\eta(\tau_2-1)\| \le \delta$; \\
			\ \ \ \ \ \ \ \ \ \ \ \ \ \ the optimal energy efficiency $\eta^*$ is obtained;\\
			\ \ \ \ \ \ \ \ \ \  \textbf{end}\\
			\ \ \ \ \ \textbf{$\pmb{\unrhd} $ end} \\
			\  4) \textbf{Output:}\\
			\ \ \ \ \ \ \ $\{\mathbf{f}_1^{*}, \mathbf{f}_2^{*}, \mathbf{w}^*\}$ and energy efficiency $\eta^*$.\\
			\bottomrule
		\end{tabular}
	\end{center}
\end{table}
\section{Simulation Results}
\label{Simulation}
In this section, simulation results are provided to verify the proposed algorithms. We consider a three-dimensional Cartesian coordinate system. The simulation settings are based on the work in \cite{irs1}, \cite{wqqirs}. 
{The locations of the base station, the Jammer, the IRS, and the legitimate user are respectively  set as $(5,0,20)$, $(5,0,15)$, $(0,100,2)$, $(3,100,0)$ and the locations of 5 different Eves are set as $(2, 105,0)$, $(2, 102.5, 0)$, $(2,100,0)$, $(2,97.5,0)$, $(2,95,0)$, respectively \cite{wqqirs}.}
The channels are generated by the model $h_{i,j}=\sqrt{G_0d_{i,j}^{-c_{i,j}}}g_{i,j}$, where $G_0=-30$ dB is the path loss at the reference point. $d_{i,j}$, $c_{i,j}$ and $g_{i,j}$ denote the distance, path loss exponent, and fading between $i$ and $j$, respectively, where $i \in \{B, J, I\}$ and $j \in \{U, (E,k)\}$ \cite{cwh5}. The path loss exponents are set as $c_{B,U}=c_{B,E,k}=c_{J,U}=c_{J,E,k}=5$, $c_{B,J}=c_{G,J}=3.5$, $c_{J,U}=2$, and $c_{J,E,k}=3$. 
{ We consider that the vertical location of the IRS is higher than those of the user and Eves. In this case, a less scattered environment is expected and one has $ c_{B,I} \le c_{B,i}$, $ c_{J,I} \le c_{J,i}$, $ i \in\{B,(E,k)\}$. For the path loss exponents between IRS and the receivers, since IRS is deployed to support the legitimate user, it is assumed that the path loss between IRS and user is smaller than that of Eves, one has $c_{I,U} \le c_{I,E,k}$.}
The bandwidth $B$ is normalized to $1$. The other parameters are set as $\xi_{I,E,k}=\xi_{J,E,k}=10^{-4}$, $P_{1,max}=P_{2, max}=P_{max}$, $\sigma^2= -105$ dBm, $\zeta=1$, $P_{BS}=P_{G}=23$ dBm, $P_{IRS}=20$ dBm, { and $\delta=10^{-7}$}.
\begin{figure}[h]
		\setlength{\abovedisplayskip}{3pt}
	\setlength{\belowdisplayskip}{3pt}
	\centering
	\begin{minipage}[t]{0.48\textwidth}
		\centering
	\includegraphics[width=3.0in]{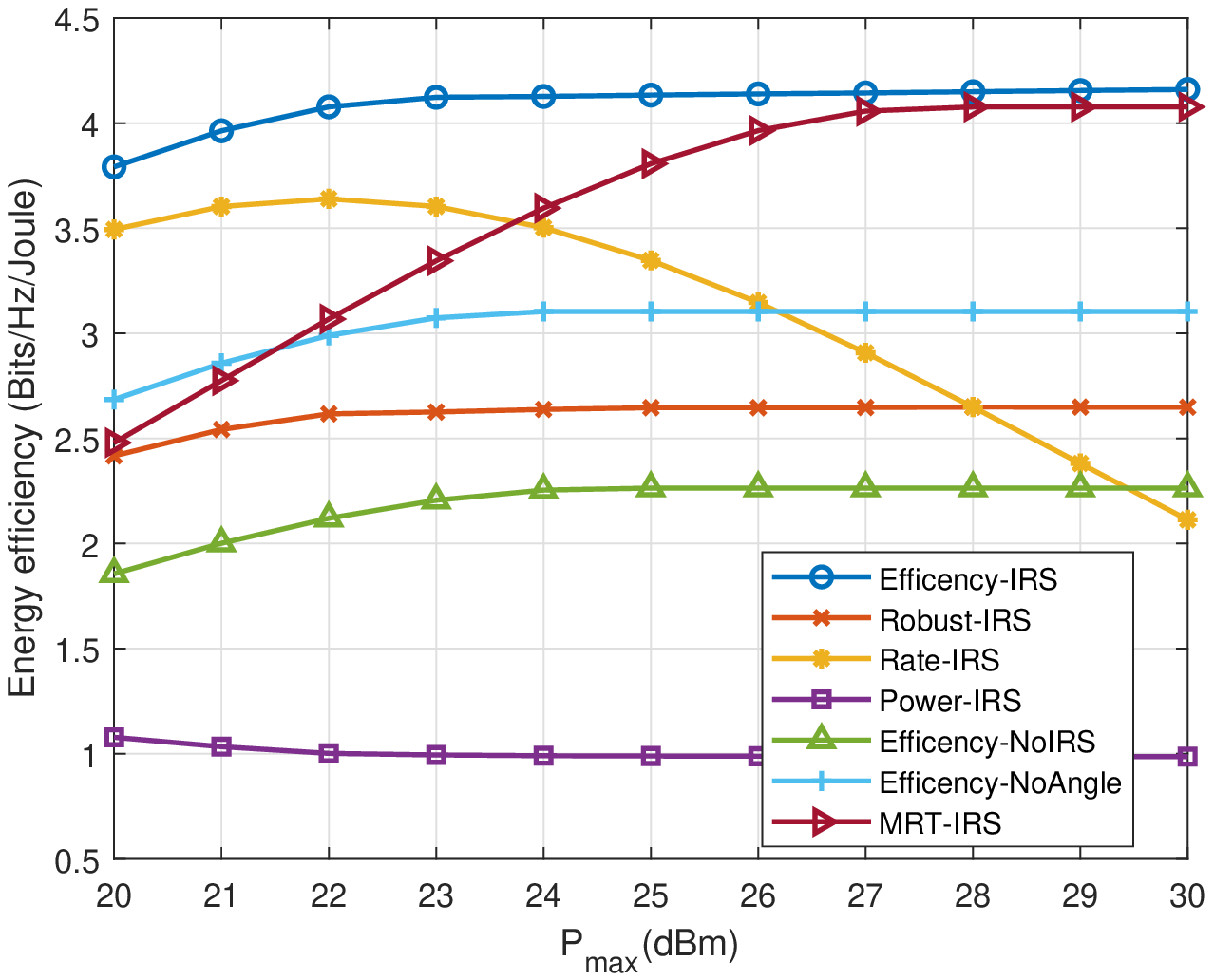}	\label{eta1}
\caption{Energy efficiency versus the maximum transmit power.}
	\end{minipage}
	\begin{minipage}[t]{0.48\textwidth}
	\centering
	\includegraphics[width=3.0in]{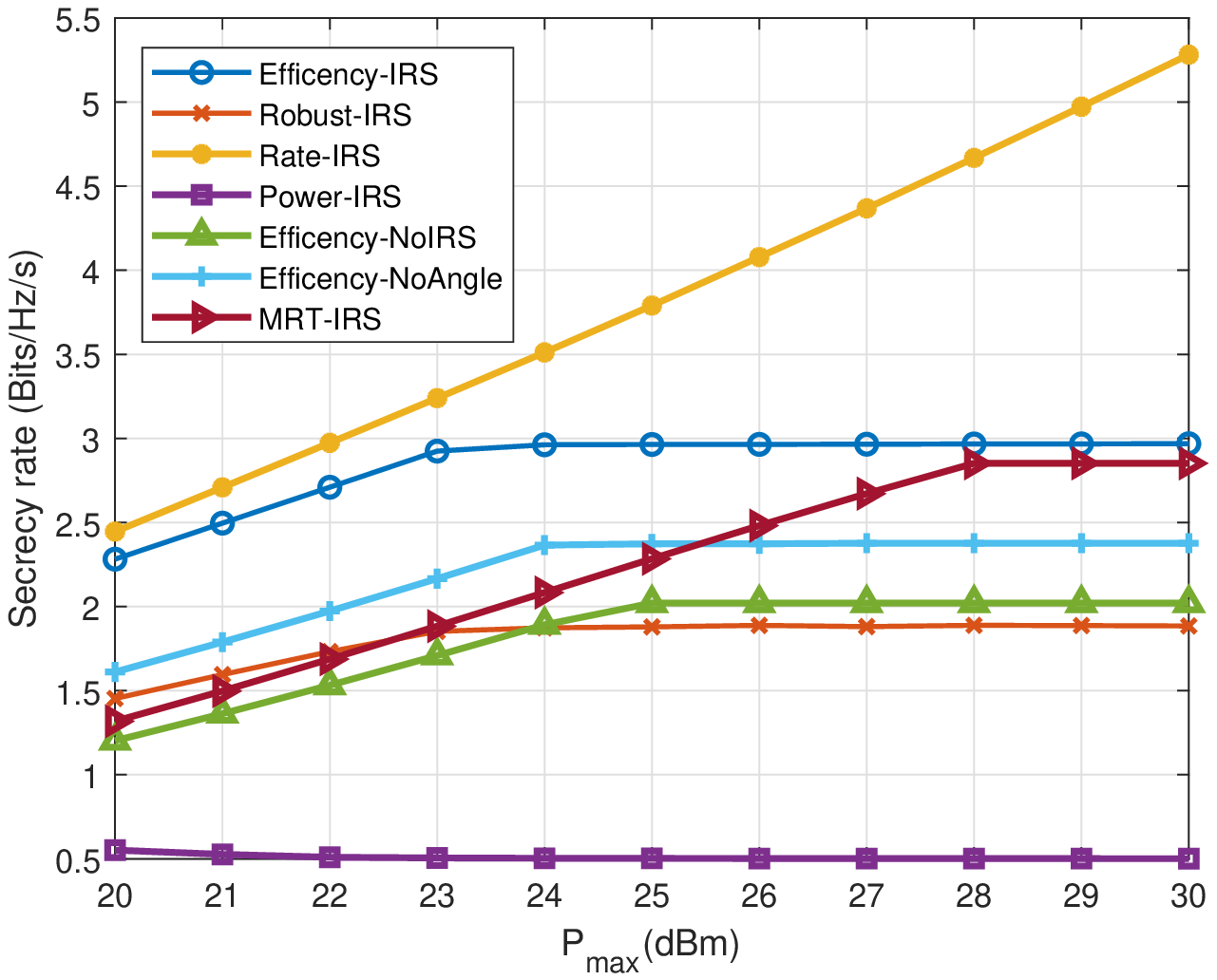}	\label{rate1}
\caption{Secrecy rate versus the maximum transmit power.}
\end{minipage}
\end{figure}

Our proposed scheme for the perfect CSI model is marked as `Efficiency-IRS'. The proposed scheme for the imperfect CSI model is marked as `Robust-IRS'. {We consider five cases as benchmarks to compare with our proposed method. The first benchmark optimizes the transmit rate, which is marked as `Rate-IRS'. The second benchmark minimizes the transmit power, which is marked as `Power-IRS'. The third benchmark without IRS is marked as `Efficiency-NoIRS'. The fourth benchmark is the method that has IRS but no phase adjustment, which is marked as `Efficiency-NoAngle'.
	The fifth benchmark is the method that is based on the maximum ratio transmission (MRT) method for beamforming design under perfect CSI case \cite{re11} and is marked `MRT-IRS'.
}

Fig. 2 shows the energy efficiency versus the maximum transmit power achieved by different designs. The minimum secrecy rate threshold is set as $R_{th}=0.5$ Bits/Hz/s. It is observed that the energy efficiency achieved by the proposed method with the perfect CSI is the best among all the schemes. 
This indicates that our proposed IRS assisted cooperative jamming scheme is efficient in improving energy efficiency and achieving secure communications. 
{
The system energy efficiency of the proposed method under the imperfect CSI condition is smaller than those achieved with the `Efficiency-IRS' method, `MRT-IRS' method, `Efficiency-Noangle' method, and `Rate-IRS' method at the beginning. This is because even without the phase optimization, IRS can help to increase the energy efficiency with the perfect CSI.}
Under the imperfect CSI, the energy efficiency degrades compared to that achieved under the perfect CSI case due to the CSI uncertainty.  
However, compared with the method without IRS, the `Robust-IRS' method can still achieve a higher energy efficiency.
This further indicates that the application of IRS is effective to improve energy efficiency even under the imperfect CSI.

It is worth noting that the system efficiencies obtained by the proposed method, the benchmark `Efficiency-NoIRS', and `Efficiency-NoAngle' all increase first with $P_{\max}$ and finally converge. 
For these methods, when the available power is limited, the increase of the secrecy rate is beneficial for the system to obtain a higher energy efficiency with only a slightly more power consumption. However, when the power availability is sufficient, e.g., $P_{\max}$ is larger than $23$ dBm in this setting, further increase of the secrecy rate causes repaid elevation of the energy consumption, which leads to a decrease in energy efficiency.
Similarly, the energy efficiency of the `Rate-IRS' method first increases with the transmit power and then gradually decreases. The reason is that this method aims to maximize the secrecy rate without the constraint on the power consumption. Thus the study shows that there is a tradeoff between the energy efficiency and the secrecy rate. The energy efficiency of the `Power-IRS' method first slightly decreases and  then keeps at a low level. The reason is that this method aims to minimize power consumption, and thus it achieves the minimum secrecy rate $R_{th}$ to save energy. In this case, both the energy efficiency and secrecy rate are relatively low.
{ The energy efficiency of the `MRT-IRS' method keeps increasing with $P_{\max}$ until reaching the highest efficiency, which is lower than that obtained with `Efficiency-IRS' method. This validates the superiority of the proposed design.}

Fig. 3 shows the achievable secrecy rate versus the maximum available transmits power $P_{\max}$. 
The secrecy rate obtained by the proposed scheme is comparable with the `Rate-IRS’ scheme when $P_{\max}$ is smaller than $23$ dBm. 
When $P_{\max}$ is larger than $23$ dBm, the `Rate-IRS' method continues to use all the available energy to increase the achievable secrecy rate.
{ The `MRT-IRS' method shows a similar trend with the `Efficiency-IRS' method but achieves  a lower rate at the optimal level, which validates the observation in Fig. 2.}
However, the proposed scheme maintains the secrecy rate at a stable level in order to achieve the maximum energy efficiency. 
Similar trends can also be observed from the `Robust-IRS' method, `Efficiency-NoIRS’ method, and `Efficiency-NoAngle’ method.
The secrecy rate achieved by the `Power-IRS' method first decreases and then stabilizes at the lowest level in order to save energy.
The achievable secrecy rate of `Robust-IRS' stabilizes at a lower level than other methods because based on the estimated channel quality, this algorithm needs to decrease the transmission rate to achieve the optimal energy efficiency under this setting.
The curves in Fig. 3 indicate that with the aided IRS, our proposed method with the perfect CSI can achieve a higher secrecy rate and obtain the maximum energy efficiency.
\begin{figure}[h]
		\setlength{\abovedisplayskip}{3pt}
	\setlength{\belowdisplayskip}{3pt}
	\centering
	\includegraphics[width=3.0in]{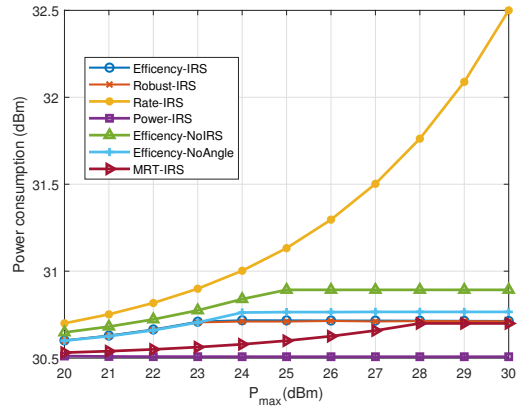}	\label{power1}
	\caption{Power consumption versus the maximum transmit power.}
\end{figure} 

Fig. 4 presents the power consumption for different methods versus $P_{\max}$. The results of all the methods in Fig. 4 are consistent with what have been shown in Fig. 2 and Fig. 3. It is worth noting that the power consumption by the proposed method with the perfect CSI and imperfect CSI are almost the same and both are quite low. This indicates that even with channel estimation errors, the `Robust-IRS' method  can still use less energy to achieve a higher rate, which demonstrates the advantage of the exploitation of IRS in improving energy efficiency.


Fig. 5 shows the energy efficiency versus the minimum secrecy rate threshold $R_{th}$. The maximum available transmit power is set to $P_{\max}=36$ dBm. 
The energy efficiency achieved by the proposed method is the best among all the schemes.  
This indicates that the IRS assisted cooperative jamming can help guarantee the secrecy rate requirement and achieve the maximum energy efficiency. 
The energy efficiency of the proposed method, and the `Efficiency-NoAngle' method initially maintain at a stable level and then decreases with the increase of $R_{th}$.
When the minimum secrecy requirement is low, a higher rate can help the system to obtain a higher energy efficiency. 
However, when $R_{th}$ is larger than the optimal rate, the system has to consume excessive energy to increase the secrecy rate in order to meet the minimum secrecy rate constraint, which causes the decrease of the energy efficiency.

{
In Fig. 5, the curves of the `Efficiency-NoIRS' method, the `Robust-IRS' method, the `MRT-IRS' method, and the `Efficiency-NoAngle' method vanish when $R_{th}$ is larger than $1.5$ Bits/Hz/s, $2$ Bits/Hz/s, $2.5$ Bits/Hz/s, and $3$ Bits/Hz/s, respectively. 
The reason is that there is no feasible solution that can meet a higher $R_{th}$ in those regions even with the maximum available transmit power.}
Moreover, the energy efficiency of the `Power-IRS' method first increases and the curve starts to decrease  when $R_{th}$ is larger than $3$ Bits/Hz/s, 
When the secrecy rate is smaller than $3$ Bits/Hz/s, the increase of the secrecy rate can bring more performance gains (say rate gain) than the energy consumption. Thus, it results in the increase of the system energy efficiency. However, when the secrecy rate becomes larger and larger, the power cost for increasing the secrecy rate goes higher than the benefits that it brings to the system, which causes a lower energy efficiency. 
This also indicates that there is a tradeoff between energy efficiency and the secrecy rate. 
The energy efficiency of `Rate-IRS' stays at a constant level. 
This can be explained by the fact that the system uses all the available power to maximize the secrecy rate without considering the achievable energy efficiency. Thus, the curve does not change with the increase of $R_{th}$.
\begin{figure}[h]
		\setlength{\abovedisplayskip}{3pt}
	\setlength{\belowdisplayskip}{3pt}
	\centering
	\begin{minipage}[t]{0.48\textwidth}
		\centering
		\includegraphics[width=3.0in]{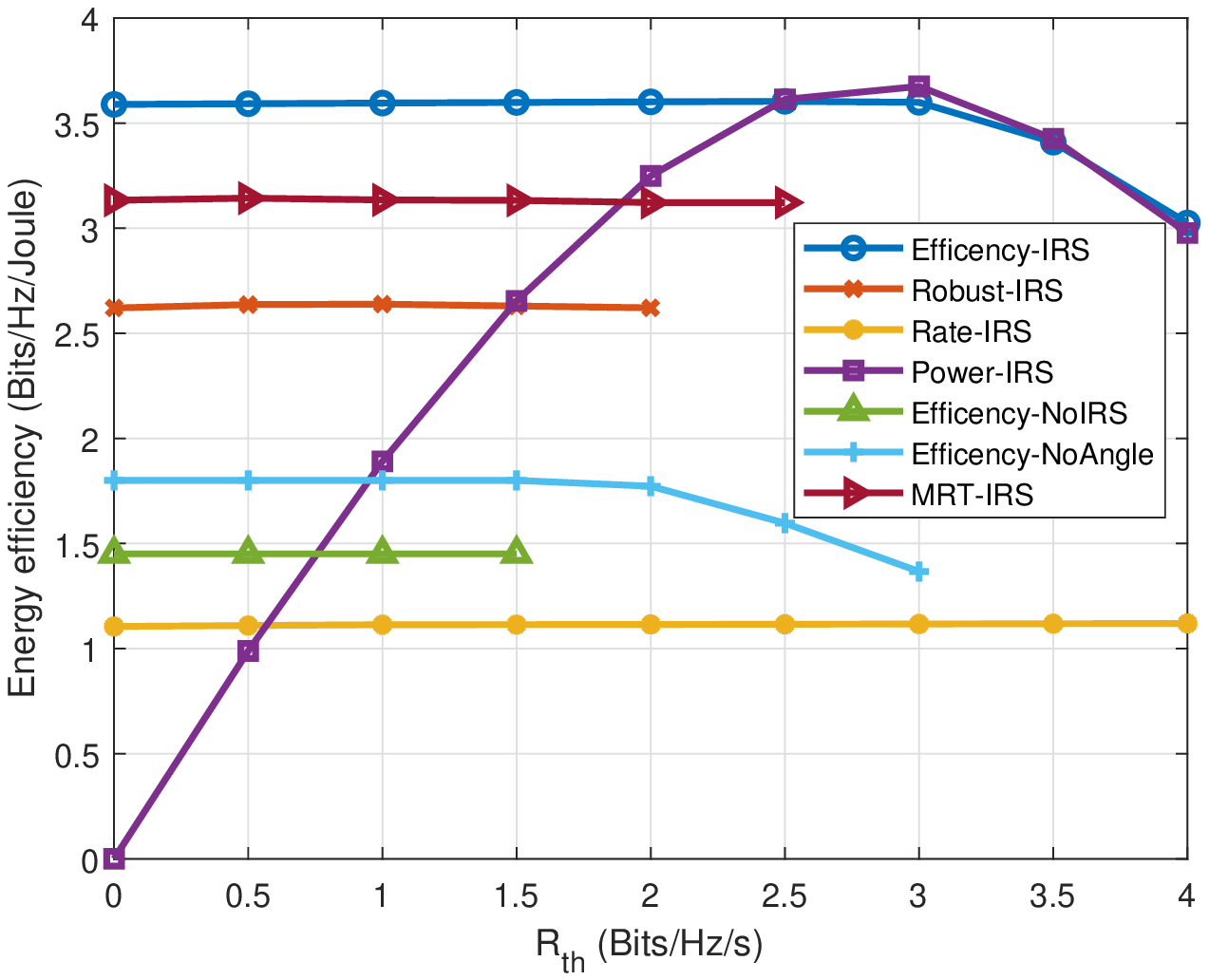}	\label{eta2}
		\caption{Energy efficiency versus the secrecy rate threshold.}
	\end{minipage}
	\begin{minipage}[t]{0.48\textwidth}
		\centering
		\includegraphics[width=3.0in]{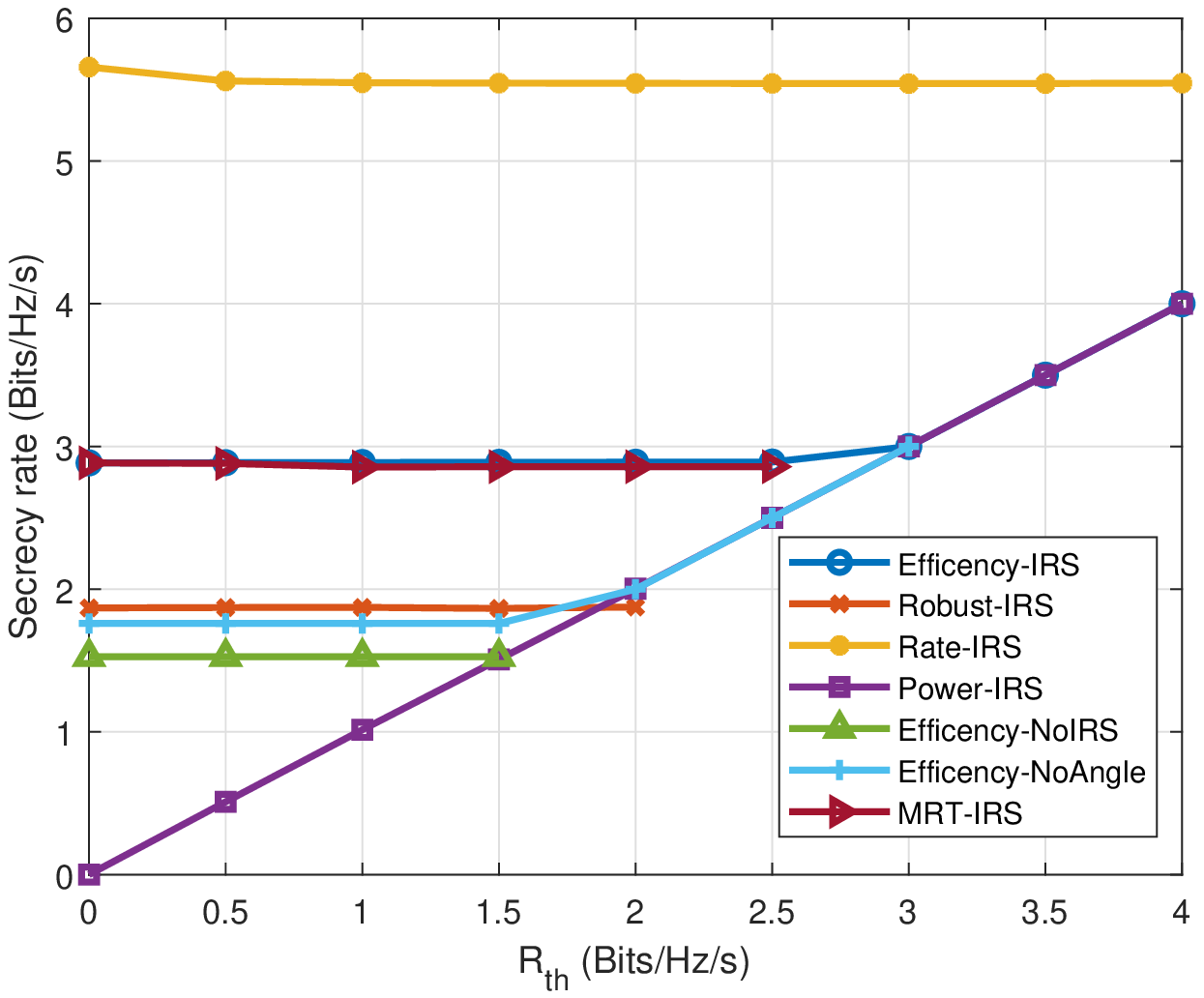}	\label{rate2}
		\caption{Achievable secrecy rate versus the secrecy rate threshold.}
	\end{minipage}
\end{figure}

A comparison of the achievable secrecy rate versus the rate threshold $R_{th}$ is presented in Fig. 6. 
The secrecy rates obtained by the proposed method, the `Efficiency-NoIRS' method, and the `Efficiency-NoAngle' method are first maintained at the stable level to guarantee the maximum energy efficiency. 
After $R_{th}$ is larger than the optimal rate, the secrecy rate constraint enforces a linear increase of the rate with $R_{th}$. Similar to the reason for Fig. 5, the missing points are caused by lack of feasible solutions for the two benchmark schemes in certain $R_{th}$ regions. 
With the assistance of the IRS, the system can use a smaller transmit power to achieve a higher secrecy rate.
Additionally, the secrecy rate of the `Power-IRS' method increases with the $R_{th}$ linearly, which also verifies the observation in Fig. 5. 
For the 'Robust-IRS method, the system efficiency and secrecy rate are both higher than those of the `Efficiency-NoIRS' method and the `Efficiency-NoAngle' method under this setting, which indicates that even with imperfect CSI, the proposed method can still achieve a better performance than the method without IRS under perfect CSI.
The secrecy rate achieved by the `Rate-IRS' method is the largest among all the methods and remains constant. 

\begin{figure}[h]
		\setlength{\abovedisplayskip}{3pt}
	\setlength{\belowdisplayskip}{3pt}
	\centering
	\begin{minipage}[t]{0.48\textwidth}
		\centering
	\includegraphics[width=3.0in]{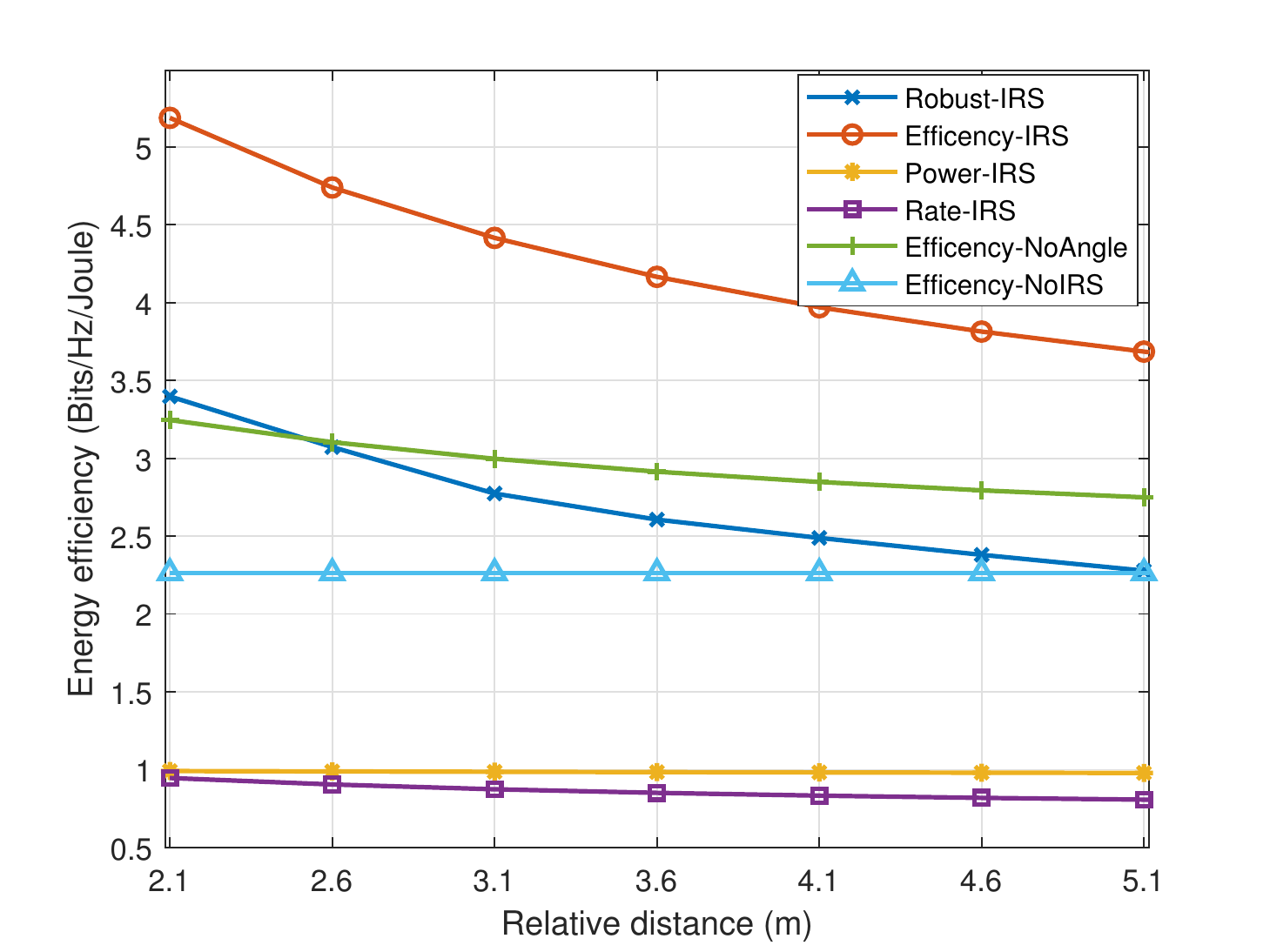}	\label{deta1}
	\caption{Energy efficiency versus the relative distance of UE-IRS.}
	\end{minipage}
	\begin{minipage}[t]{0.48\textwidth}
		\centering
		\includegraphics[width=3.0in]{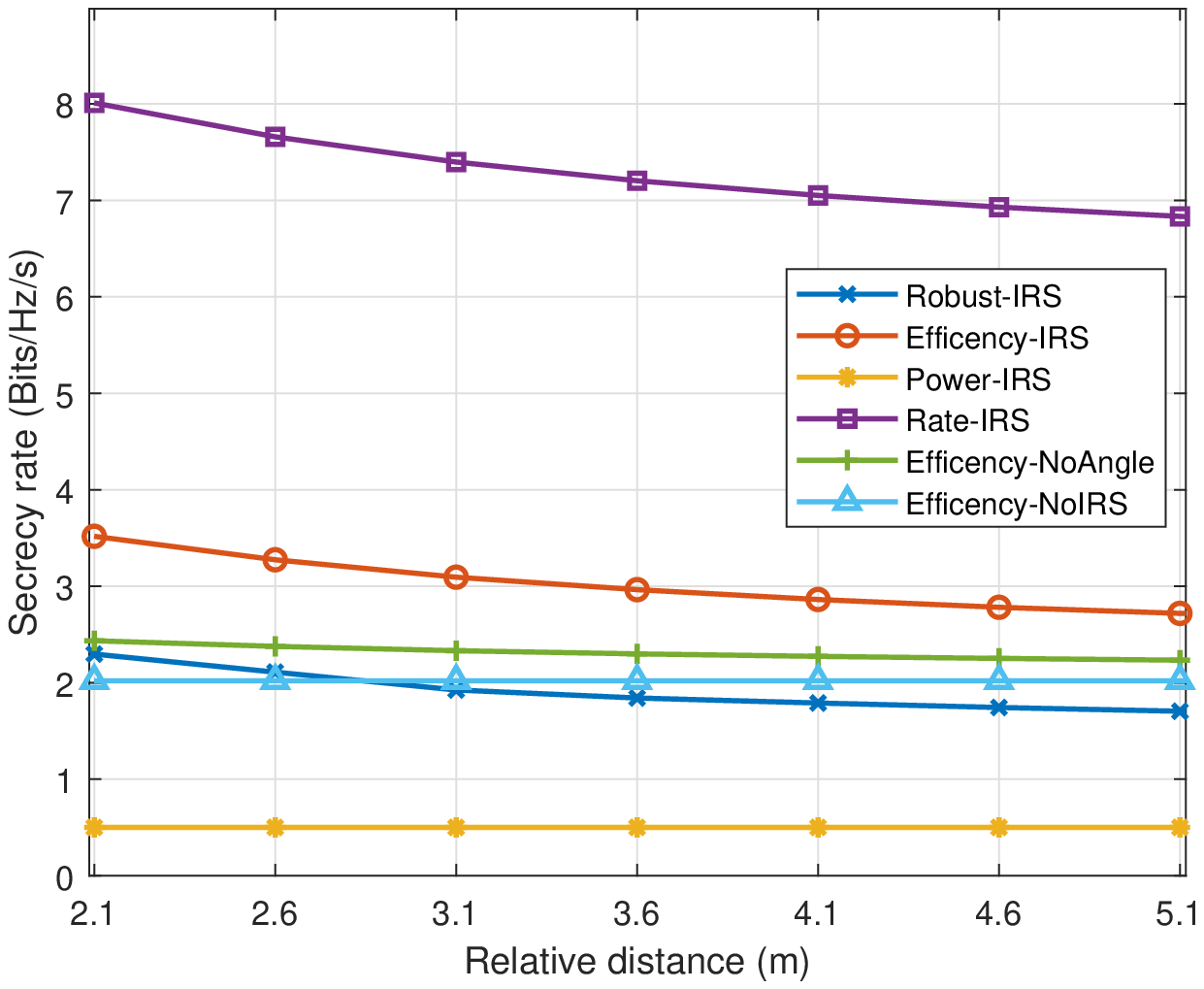}	\label{drate1}
		\caption{Secrecy rate versus the relative distance of UE-IRS.}
	\end{minipage}
\end{figure}

{
Fig. 7 shows the energy efficiency versus the relative distance between the user and IRS. The curves for all the methods with IRS decrease  with the increase of the distance. This is because the increase of the distance results in the increase of the path loss and the reduction of the power gain from the reflecting path through the IRS. Therefore, the achievable secure rate and energy efficiency both are decreased. It is also seen that the `Efficiency-IRS' method still has the highest performance among all the methods, which validates the superiority of our proposed design.

Fig. 8 shows the achievable secure rate versus the relative distance. The trend is consistent with that shown in Fig. 7. It is worth noting that although the secrecy rate of the `Robust-IRS' method is lower than that obtained with `Efficiency-NoIRS' method due to the uncertainty under the imperfect CSI, the energy efficiency of the `Robust-IRS' is still larger than that achieved with the `Efficiency-NoIRS' method. This further demonstrates the efficiency of the proposed robust design.
}

\begin{figure}[h]
		\setlength{\abovedisplayskip}{3pt}
	\setlength{\belowdisplayskip}{3pt}
	\centering
	\begin{minipage}[t]{0.48\textwidth}
		\centering
			\includegraphics[width=3.0in]{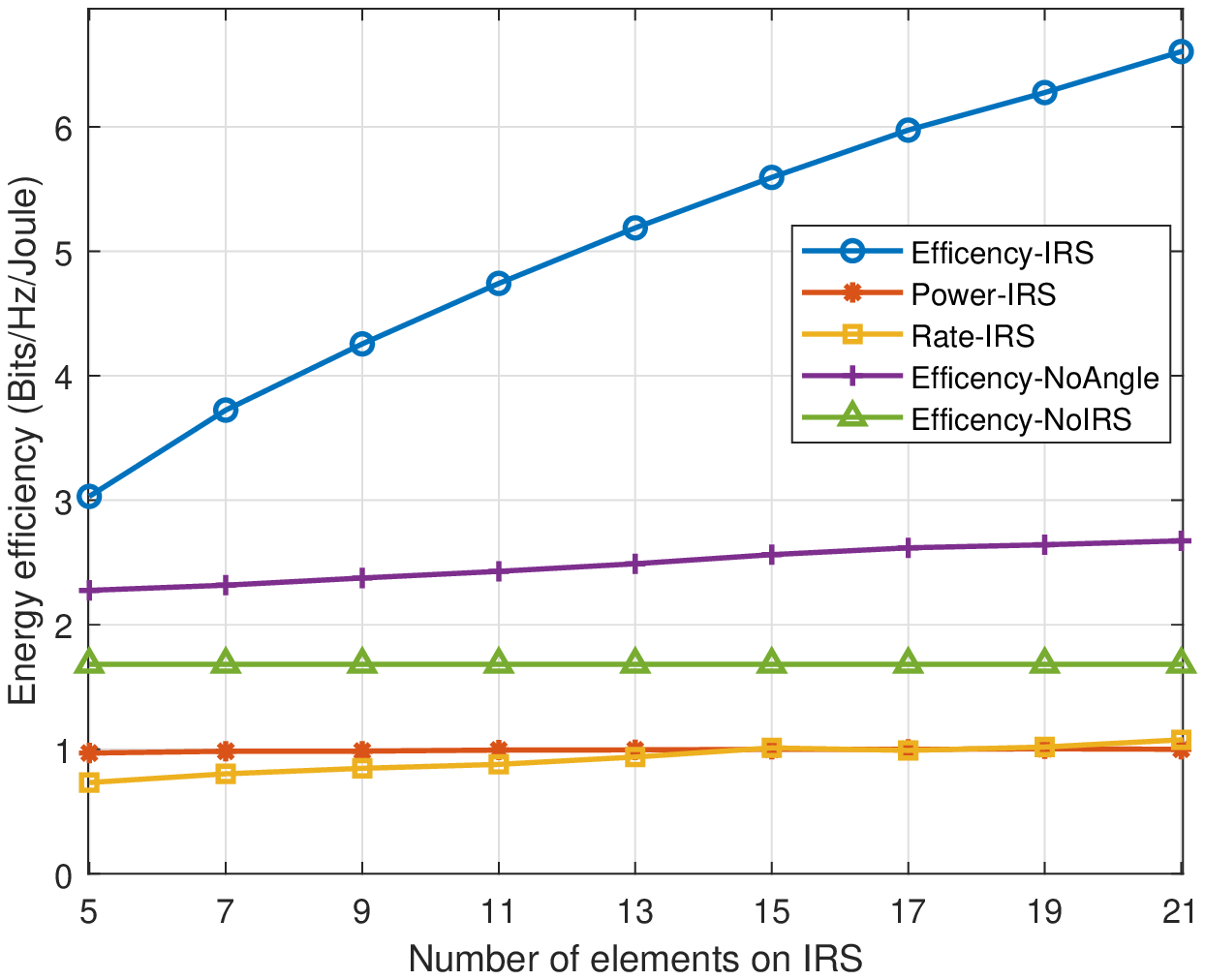}	\label{eleta}
		\caption{Energy efficiency versus the number of elements on IRS.}
	\end{minipage}
	\begin{minipage}[t]{0.48\textwidth}
		\centering
		\includegraphics[width=3.0in]{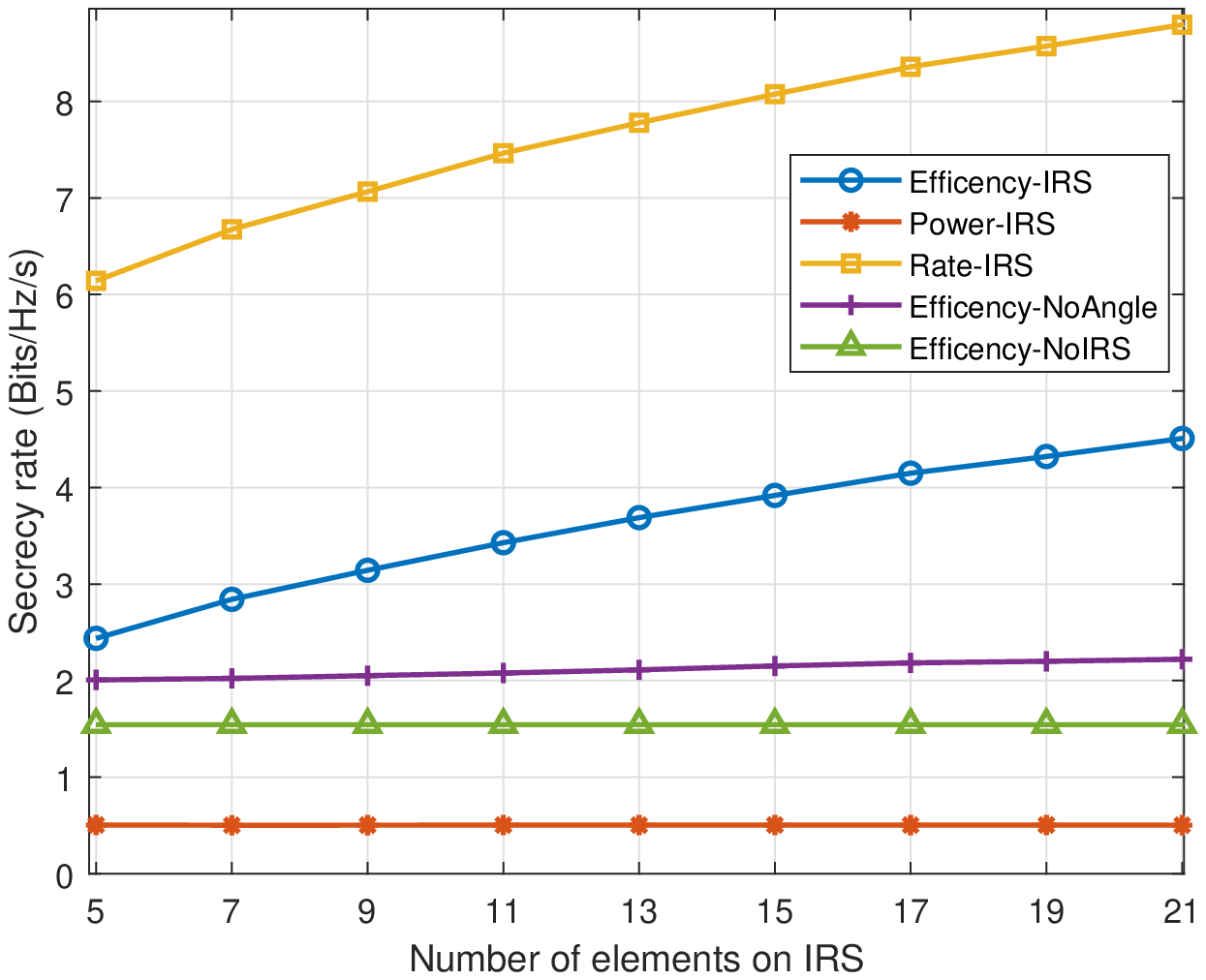}	\label{elrate}
		\caption{Secrecy rate versus the number of elements on IRS.}
	\end{minipage}
\end{figure}

{
Fig. 9 shows energy efficiency versus the number of reflecting elements on the IRS. It is seen that the higher the number of the reflecting elements on the IRS, the better the energy efficiency obtained in the IRS- assisted network.
The reason is that a better performance can be achieved by employing a higher number of reflecting elements to enhance the desired signals for the legitimate user.
The increasing gain of the proposed method is higher than those obtained with other benchmarks, which demonstrates that the proposed scheme can effectively exploit the IRS to assist the secure transmission.

Fig. 10 shows the impact of the number of IRS elements on the achievable secure rate. It can be observed that both the `Rate-IRS' method and the `Efficiency-IRS' method achieve evident improvement on the secure rate. Compared with Fig. 9, it is worth noting that energy efficiency of the `Rate-IRS' method is also increased. 
The reason is that the diversity gain can be achieved by increasing the number of the reflecting elements.
}
\section{Conclusions}
In this paper, an IRS-assisted MISO wireless communication network was considered with the independently cooperative jamming in order to achieve secure communications. The energy efficiency was maximized by jointly optimizing the beamforming, jamming precode vectors, and IRS phase shift matrix under both perfect and imperfect CSI conditions. Two alternating algorithms were proposed to solve the challenging non-convex fractional optimization problems. It was shown that our proposed method outperforms other schemes in terms of energy efficiency. Although there is a tradeoff between the secrecy rate and energy efficiency, the application of IRS can effectively improve the energy efficiency even under the imperfect CSI case. 
{
The proposed alternating algorithm can be extended to the multi-IRS multi-UE MIMO communication network and the research will be done in our future works.}

\begin{IEEEbiography}[{\includegraphics[width=1in,height=1.25in,clip,keepaspectratio]{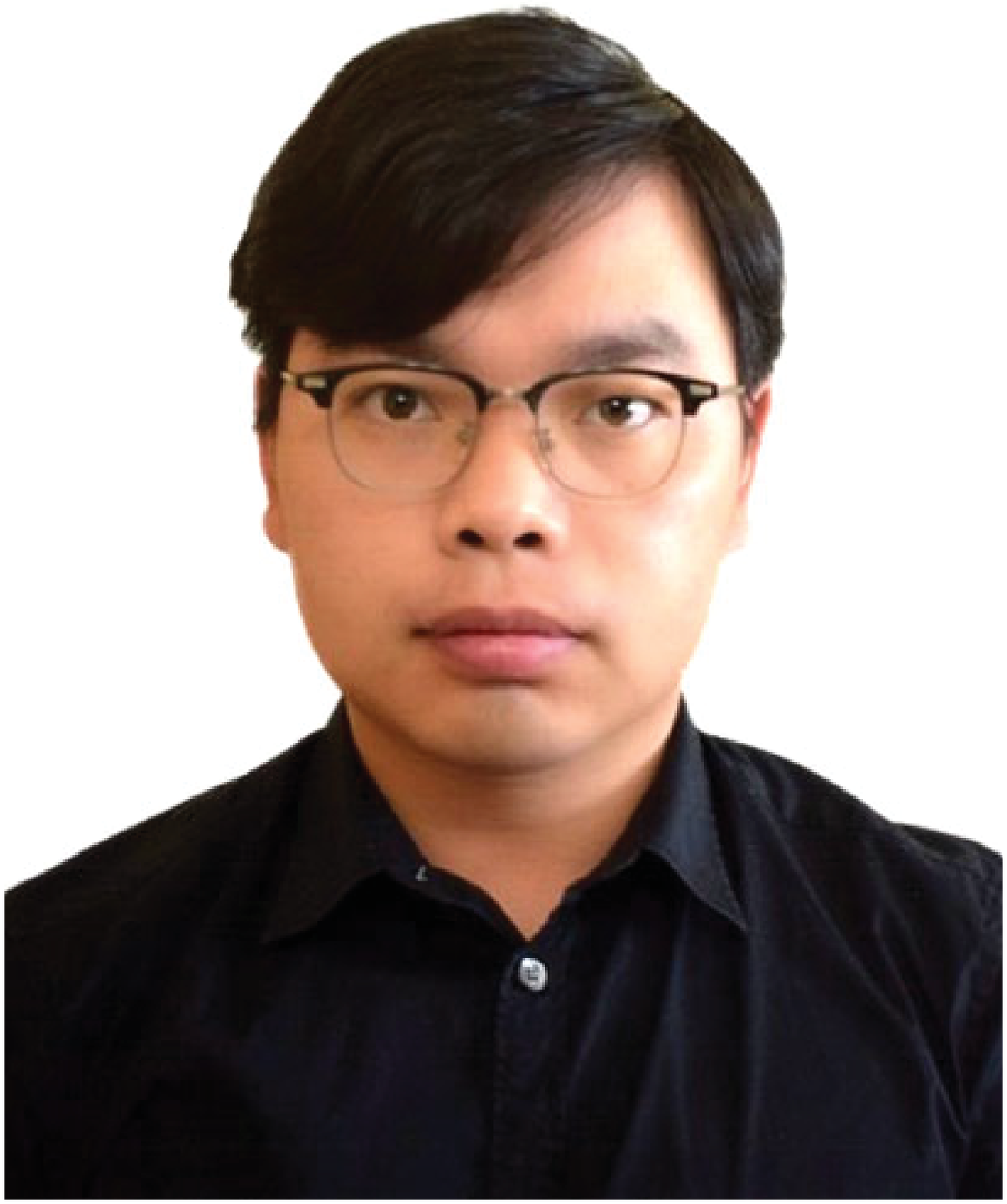}}]{Qun Wang (S'16)} received the M.S. degree from Xidian University, Xi’an, China, in 2016. He is currently pursuing the Ph.D. degree with the Department of Electrical and Computer Engineering, Utah State University, Logan, UT, USA. His research interests include mobile edge computing, nonorthogonal multiple access, intelligent reflect surface, MIMO, and machine learning.
\end{IEEEbiography}
\begin{IEEEbiography}[{\includegraphics[width=1in,height=1.25in,clip,keepaspectratio]{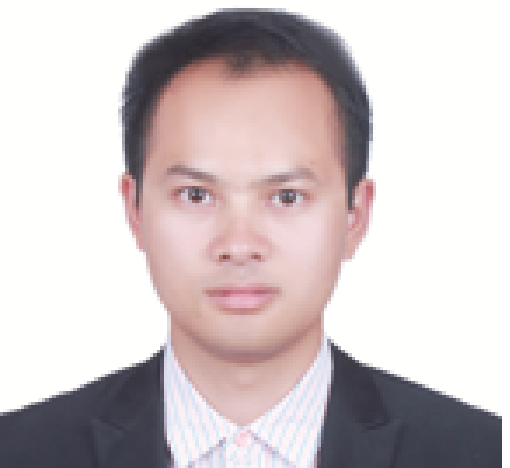}}]{Fuhui Zhou (M’16-SM’20)} is currently a Full Professor with the College of Electronic and Information Engineering, Nanjing University of Aeronautics and Astronautics. He has worked as a Senior Research Fellow at Utah State University. He received the Ph. D. degree from Xidian University, Xi’an, China, in 2016. His research interests focus on cognitive radio, edge computing, machine learning, and resource allocation. He has published more than 100 papers, including IEEE Journal of Selected Areas in Communications, IEEE Transactions on Wireless Communications, IEEE Transactions on Communications, IEEE Wireless Communications, IEEE Network, IEEE GLOBECOM, etc. He was awarded as Young Elite Scientist Award of China. He has served as Technical Program Committee (TPC) member for many International conferences, such as IEEE GLOBECOM, IEEE ICC, etc. He serves as an Editor of IEEE Transactions on Communications, IEEE Systems Journal, IEEE Wireless Communications Letters, IEEE Access and Physical Communications. He also serves as co-chair of IEEE ICC 2019 and IEEE Globecom 2019 workshop on “Advanced Mobile Edge /Fog Computing for 5G Mobile Networks and Beyond”.  
\end{IEEEbiography}
\begin{IEEEbiography}[{\includegraphics[width=1in,height=1.25in,clip,keepaspectratio]{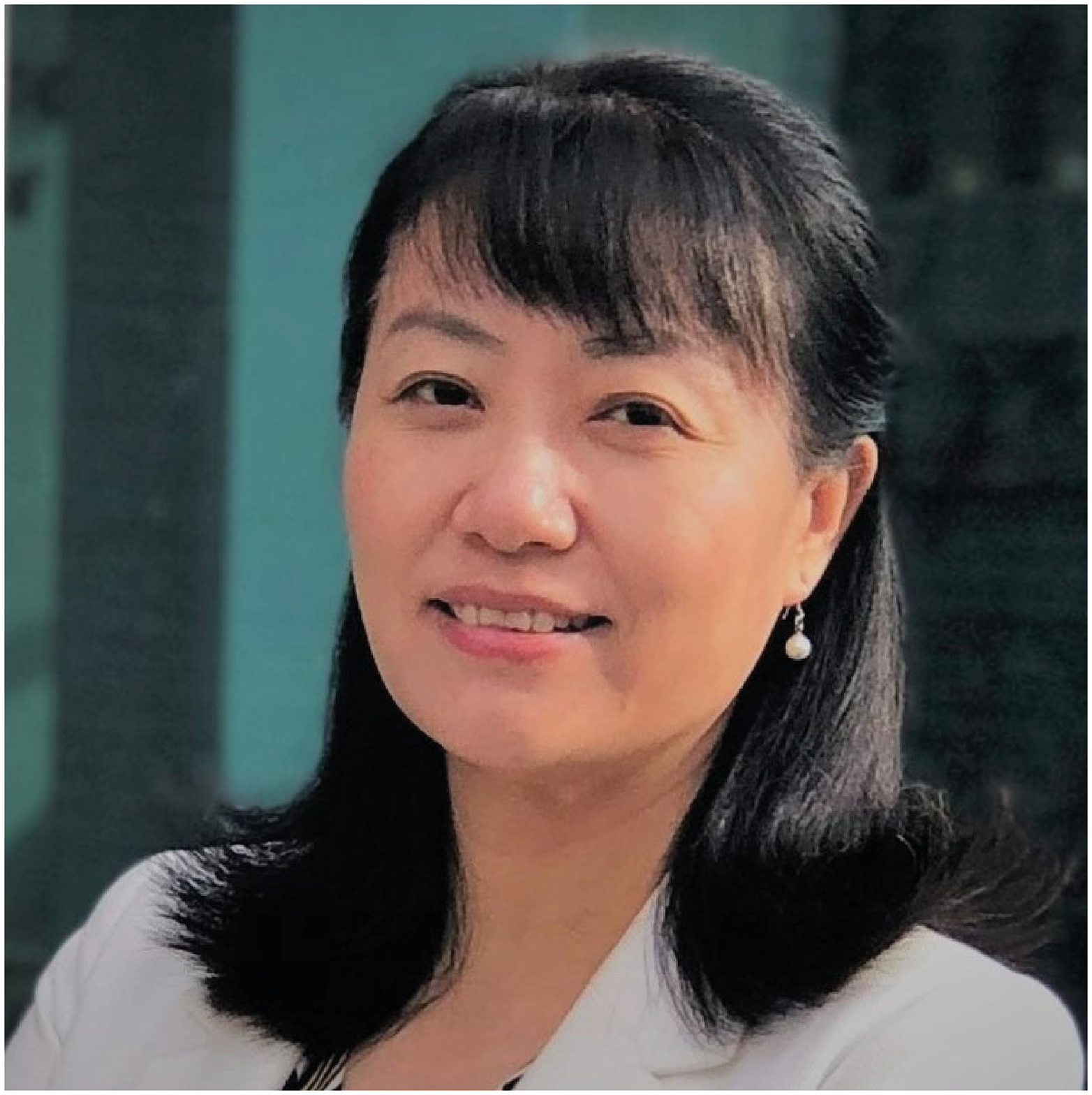}}]{Rose Qingyang Hu (S’95–M’98–SM’06-F’20)} received the B.S. degree from the University of Science and Technology of China, the M.S. degree from New York University, and the Ph.D. degree from the University of Kansas. Besides a decade academia experience, she has more than 10 years of R\&D experience with Nortel, Blackberry, and Intel as a Technical Manager, a Senior Wireless System Architect, and a Senior Research Scientist, actively participating in industrial 3 G/4 G technology development, standardization, system level simulation, and performance evaluation. 
	She is a Professor with the Electrical and Computer Engineering Department and Associate Dean for research of College of Engineering at Utah State University. She also directs Communications Network Innovation Lab at Utah State University. Her current research interests include next-generation wireless system design and optimization, Internet of Things, Cyber Physical system, Mobile Edge Computing, V2X communications, artificial intelligence in wireless networks, wireless system modeling and performance analysis. She has published extensively in top IEEE journals and conferences and also holds numerous patents in her research areas. She is currently serving on the editorial boards of the IEEE TRANSACTIONS ON WIRELESS COMMUNICATIONS, IEEE TRANSACTIONS ON VEHICULAR TECHNOLOGY, IEEE Communications Magazine and IEEE WIRELESS COMMUNICATIONS. She also served as the TPC Co-Chair for the IEEE ICC 2018. She is an IEEE Communications Society Distinguished Lecturer Class 2015-2018 and a recipient of prestigious Best Paper Awards from the IEEE GLOBECOM 2012, the IEEE ICC 2015, the IEEE VTC Spring 2016, and the IEEE ICC 2016. She is member of Phi Kappa Phi Honor Society.
\end{IEEEbiography}
\begin{IEEEbiography}[{\includegraphics[width=1in,height=1.25in,clip,keepaspectratio]{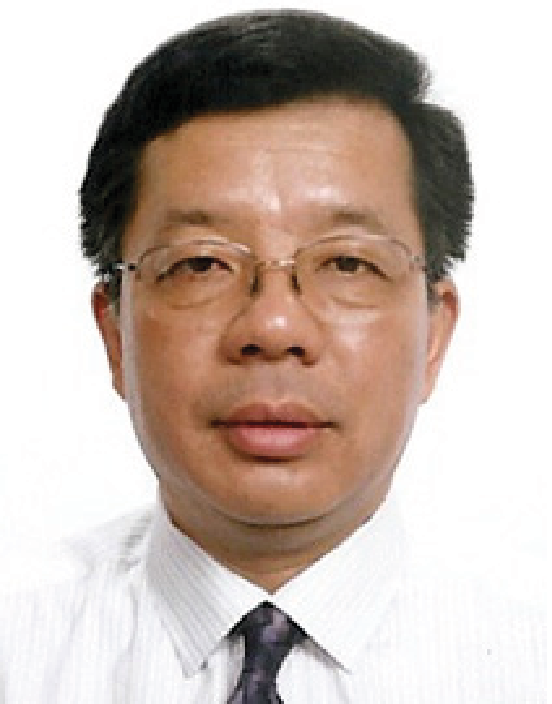}}]{Yi Qian (M’95-SM’07-F’19) }  received the Ph.D. degree in electrical engineering from Clemson University. He is a Professor with the Department of Electrical and Computer Engineering, University of Nebraska-Lincoln (UNL). His research interests include communications and systems, and information and communication network security. He was previously Chair of the IEEE Technical Committee for Communications and Information Security. He was the Technical Program Chair for IEEE International Conference on Communications 2018. He serves on the editorial boards of several international journals and magazines, including as the Editor-in-Chief for IEEE Wireless Communications. He was a Distinguished Lecturer for IEEE Vehicular Technology Society. He is currently a Distinguished Lecturer for IEEE Communications Society.
\end{IEEEbiography}
\end{document}